\documentclass[a4paper, 12pt]{article}

\pdfoutput = 1

\usepackage{styleBuding}
\usepackage{bm}
\usepackage{color,listings}
\usepackage[usenames,dvipsnames,svgnames,table]{xcolor}
\usepackage{amsthm, amsmath}
\usepackage{amssymb}
\usepackage{mathrsfs}
\usepackage{graphicx}
\usepackage{slashed}
\usepackage{soul}
\usepackage{multirow}
\usepackage{subfigure}
\usepackage{diagbox}
\usepackage{siunitx} % for physics units
\usepackage{url} %for link in @article
\usepackage[utf8]{inputenc}
\usepackage[figuresright]{rotating}
\definecolor{back}{HTML}{F8F8F8}
\usepackage{epstopdf}
\usepackage{lipsum}
\newcommand\blfootnote[1]{%
	\begingroup
	\renewcommand\thefootnote{}\footnote{#1}%
	\addtocounter{footnote}{-1}%
	\endgroup
}

\newcommand{\rom}[1]{\uppercase\expandafter{\romannumeral #1\relax}}

\allowdisplaybreaks
\DeclareUnicodeCharacter{2212}{-}

\let\jnfont=\rm
\def\NPB#1,{{\jnfont Nucl.\ Phys.\ B }{\bf #1},}
\def\PLB#1,{{\jnfont Phys.\ Lett.\ B }{\bf #1},}
\def\EPJC#1,{{\jnfont Eur.\ Phys.\ Jour.\ C }{\bf #1},}
\def\PRD#1,{{\jnfont Phys.\ Rev.\ D }{\bf #1},}
\def\PRL#1,{{\jnfont Phys.\ Rev.\ Lett.\ }{\bf #1},}
\def\MPLA#1,{{\jnfont Mod.\ Phys.\ Lett.\ A }{\bf #1},}
\def\JPG#1,{{\jnfont J.\ Phys.\ G}{\bf #1},}
\def\CTP#1,{{\jnfont Commun.\ Theor.\ Phys.\ }{\bf #1},}
\def\ZPC#1,{{\jnfont Z.\ Phys.\ C }{\bf #1},}
\def\JHEP#1,{{\jnfont JHEP \ }{\bf #1},}

\title{Status of the singlino-dominated dark matter in general Next-to-Minimal Supersymmetric Standard Model}
\author{Junjie Cao$^{a,b}$, Xinglong Jia$^b$, Lei Meng$^b$, Yuanfang Yue$^{b,*}$\blfootnote{*Corresponding authors.}, Di Zhang$^{b,*}$}
\affiliation{ $^a$  Schools of Physics, Shandong University, Jinan, Shandong 250100, China}
\affiliation{ $^b$ Department of Physics, Henan Normal University, Xinxiang 453007, China}
\emailAdd{junjiec@alumni.itp.ac.cn}
\emailAdd{JiaXinglong1996@outlook.com}
\emailAdd{mel18@foxmail.com}
\emailAdd{yueyuanfang@htu.edu.cn}
\emailAdd{dz481655@gmail.com}

\abstract{
With the rapid progress of dark matter direct detection experiments, the attractiveness of the popular bino-dominated dark matter in economical supersymmetric theories is fading.
As an alternative, the singlino-dominated dark matter in general Next-to-Minimal Supersymmetric Standard Model (NMSSM) is paying due attention. This scenario has the following distinct characteristics: free from the tadpole problem and the domain-wall problem of the NMSSM with a $Z_3$-symmetry, predicting more stable vacuum states than the $Z_3$-NMSSM, capable of forming an economical secluded dark matter sector to yield the dark matter experimental results naturally, and readily weaken the restrictions from the LHC search for SUSY. Consequently, it can explain the muon g-2 anomaly in broad parameter space that agrees with various experimental results while simultaneously breaking the electroweak symmetry naturally. In this study, we show in detail how the scenario coincides with the experiments, such as the SUSY search at the LHC, the dark matter search by the LZ experiment, and the improved measurement of the muon g-2. We provide a simple and clear picture of the physics inherent in the general NMSSM.

}

\begin{document}
    \maketitle
    \flushbottom
\newpage
\section{\label{Introduction}Introduction}

Thanks to the rapid progress of particle physics experiments in recent years, rich information about supersymmetry (SUSY) has been accumulated. The Run-II data of the Large Hadron Collider (LHC) enable scientists to explore the properties of winos and higgsinos with their masses up to about $1060~{\rm GeV}$ for $m_{\tilde{\chi}_1^0} \lesssim 400~{\rm GeV}$ and $900~{\rm GeV}$ for $m_{\tilde{\chi}_1^0} \lesssim 240~{\rm GeV}$~\cite{ATLAS:2019lff}, respectively, where $\tilde{\chi}_1^0$ denotes the lightest neutralino acting as the lightest supersymmetric particle (LSP), and thus dark matter (DM) candidate under the assumption of $R$-parity conservation~\cite{Jungman:1995df}, and $m_{\tilde{\chi}_1^0}$ is its mass. The data also exclude with statistical methods squarks lighter than approximate $1850~{\rm GeV}$ when the LSP is massless~\cite{ATLAS:2021jyv,ATLAS:2020syg}. Further, the combined measurement of the muon anomalous magnetic moment, $a_{\mu} \equiv (g-2)_\mu/2$, by the E989 experiment at Fermilab Laboratory~\cite{Abi:2021gix} and the E821 experiment at the Brookhaven National Laboratory (BNL)~\cite{Bennett:2006fi} indicates a 4.2$\sigma$ discrepancy from the Standard Model's prediction~\cite{Aoyama:2020ynm,Aoyama:2012wk,Aoyama:2019ryr,Czarnecki:2002nt,Gnendiger:2013pva,Davier:2017zfy,Keshavarzi:2018mgv,Colangelo:2018mtw,Hoferichter:2019gzf,Davier:2019can,Keshavarzi:2019abf,Kurz:2014wya,Melnikov:2003xd,Masjuan:2017tvw,Colangelo:2017fiz,Hoferichter:2018kwz,Gerardin:2019vio,Bijnens:2019ghy,Colangelo:2019uex,Blum:2019ugy,Colangelo:2014qya}. Although this difference may be caused by the uncertainties in calculating the hadronic contribution to the moment, as suggested by the recent lattice simulation of the BMW collaboration~\cite{Borsanyi:2020mff}, it was widely conjectured to arise from new physics beyond the SM (see, e.g., Ref.~\cite{Athron:2021iuf} and the references therein). Along this direction, it is remarkable that once the difference is confirmed to originate from SUSY effects, salient features of the theory, e.g., the mass spectra of the electroweakinos and sleptons, will be revealed ~\cite{Martin:2001st,Domingo:2008bb,Moroi:1995yh,Hollik:1997vb,Athron:2015rva,Endo:2021zal,Stockinger:2006zn,Czarnecki:2001pv,Cao:2011sn,Kang:2016iok,Zhu:2016ncq,Yanagida:2017dao, Hagiwara:2017lse,Cox:2018qyi,Tran:2018kxv,Padley:2015uma,
	Choudhury:2017fuu,Okada:2016wlm,Du:2017str, Ning:2017dng, Wang:2018vxp,Yang:2018guw,Liu:2020nsm,Cao:2019evo,Cao:2021lmj,Ke:2021kgy,Lamborn:2021snt,Li:2021xmw,Nakai:2021mha,Li:2021koa,Kim:2021suj,Li:2021pnt,Altmannshofer:2021hfu,
	Baer:2021aax,Chakraborti:2021bmv,Aboubrahim:2021xfi,Iwamoto:2021aaf,Chakraborti:2021dli,Cao:2021tuh,Yin:2021mls,Zhang:2021gun,Ibe:2021cvf,
	Han:2021ify,Wang:2021bcx,Zheng:2021gug,Chakraborti:2021mbr,Aboubrahim:2021myl,Ali:2021kxa,Wang:2021lwi,Chakraborti:2020vjp,Baum:2021qzx,Cao:2022chy,Cao:2022htd,Domingo:2022pde}. In addition, the LUX-ZEPLIN (LZ) experiment just released its first results about the direct search for DM, where the sensitivities to spin-independent (SI) and spin-dependent (SD) cross-sections of DM-nucleon scattering have reached about $6.0 \times 10^{-48}~{\rm cm^2}$ and $1.0 \times 10^{-42}~{\rm cm^2}$, respectively~\cite{LZ:2022ufs}, when the DM mass ranges from $20~{\rm GeV}$ to $40~{\rm GeV}$. These unprecedent precisions limit strongly the DM couplings to the SM particles. Given that these remarkable achievements reflect different aspects of SUSY, we are motivated to survey their combined impacts on the theory.

As an economical realization of SUSY, Next-to-Minimal Supersymmetric Standard Model (NMSSM) predicts several viable DM candidates. On the premise of explaining both the experimentally measured density of the DM in the universe and the muon g-2 anomaly, the candidate must be bino- or singlino-dominated lightest neutralino~\cite{Cao:2021tuh,Cao:2022htd}\footnote{In the case that the lightest left-handed sneutrino acts as a DM candidate, its interaction with $Z$-boson predicts a much smaller density than its measured value, i.e., $\Omega \hbar^2 \ll 0.12$, and meanwhile an unacceptably large DM-nucleon scattering rate~\cite{Falk:1994es}. For wino- or higgsino-dominated DM case, the density is at the order of $10^{-3}$ by our calculation. These cases were surveyed in MSSM to explain the muon g-2 anomaly in Ref.~\cite{Chakraborti:2021kkr}.}. Concerning the former case, the cross-sections of the DM-nucleon scattering are approximated by~\cite{Baum:2017enm,Cao:2019qng}
\begin{eqnarray}
\sigma_{\tilde{\chi}_1^0-N}^{\rm SI} & \simeq & 5 \times 10^{-45} {\rm cm^2} \left ( \frac{C_{\tilde{\chi}_1^0 \tilde{\chi}_1^0 h}}{0.1} \right )^2 \left (\frac{m_h}{125 {\rm GeV}} \right )^2, \\
\sigma_{\tilde{\chi}_1^0-N}^{\rm SD} & \simeq & C_N \times \left ( \frac{C_{\tilde{\chi}_1^0 \tilde{\chi}_1^0 Z}}{0.1} \right )^2,
\end{eqnarray}
where $C_p \simeq 1.8 \times 10^{-40}~{\rm cm^2} $ for protons, $C_n \simeq 1.4 \times 10^{-40}~{\rm cm^2} $ for neutrons, and $C_{\tilde{\chi}_1^0 \tilde{\chi}_1^0 h}$ and $C_{\tilde{\chi}_1^0 \tilde{\chi}_1^0 Z}$ represent DM couplings to the SM-like Higgs boson discovered at the LHC and $Z$-boson, respectively. These couplings take the following form~\cite{Pierce:2013rda,Calibbi:2014lga,Cheung:2014lqa}
\begin{eqnarray}
C_{\tilde{\chi}_1^0 \tilde{\chi}_1^0 h} & \simeq & e \tan \theta_W \frac{m_Z}{\mu (1 - m_{\tilde{\chi}_1^0}^2/\mu^2)} \left ( \cos (\beta + \alpha) + \sin (\beta - \alpha) \frac{m_{\tilde{\chi}_1^0}}{\mu} \right )  \nonumber \\
& \simeq & e \tan \theta_W \frac{m_Z}{\mu (1 -  m_{\tilde{\chi}_1^0}^2/\mu^2)} \left ( \sin 2 \beta + \frac{m_{\tilde{\chi}_1^0}}{\mu} \right ), \\
C_{\tilde{\chi}_1^0 \tilde{\chi}_1^0 Z} & \simeq & \frac{e \tan \theta_W \cos 2 \beta}{2} \frac{m_Z^2}{\mu^2 - m_{\tilde{\chi}_1^0}^2},
\end{eqnarray}
where $\theta_W$ is the weak mixing angle, $ m_{\tilde{\chi}_1^0}$ denotes the lightest neutralino mass that relates with the bino mass $M_1$ by $ m_{\tilde{\chi}_1^0} \simeq M_1$, $\mu$ represents higgsino mass, $\tan \beta = v_u/v_d$ is the ratio of Higgs vacuum expectation values, and $\alpha$ is the mixing angle of the CP-even Higgs states.
The formulae indicate that the DM direct detection experiments alone can set a lower bound on the magnitude of $\mu$. In Ref.~\cite{Cao:2022htd}, we performed a sophisticated scan over the parameter space of the NMSSM with a discrete $Z_3$-symmetry ($Z_3$-NMSSM) to explain the muon g-2 anomaly. During this process, we considered the constraints from the DM direct search experiment XENON-1T~\cite{Aprile:2018dbl,Aprile:2019dbj} and the LHC search for SUSY. We found that the XENON-1T experiment had set the limit $\mu \gtrsim 300~{\rm GeV}$, and the LHC constraints could further improve it to about $450~{\rm GeV}$. In addition, if one replaces the XENON-1T restrictions with the LZ restrictions to reanalyze the samples obtained in the scan, the bounds become about $380~{\rm GeV}$ and $550~{\rm GeV}$, respectively. We emphasized that the enhanced bounds on $\mu$ have at least two important implications. One is that it deteriorates the naturalness in the electroweak symmetry breaking, which is reflected by the need for more fine-tuned cancellations to obtain $Z$-boson mass~\cite{Baer:2012uy}. The other is that the heavy higgsinos prefer moderately light gauginos and/or sleptons to explain the muon g-2 anomaly, which can significantly strengthen the LHC restrictions. Specifically, the Bayesian evidence obtained in Ref.~\cite{Cao:2022htd} is reduced by a factor of about 6 when the LZ experiment is used to limit the $Z_3$-NMSSM. This result reflects an intrinsic tension between the improving DM direct search experiments and the LHC experiments in explaining the anomaly. Consequently, the bino-dominated DM case is becoming tightly limited, given the smooth advancement of the DM direct detection experiments\footnote{Recently, three authors of this work, namely X. Jia, L. Meng, and Y. Yue, surveyed the Minimal Supersymmetric Standard Model (MSSM) similarly to this study. They took great pains to explore the theory's parameter space in detail and confirmed the above conclusions.}. This feature makes it less attractive.

The singlino-dominated DM differs from the bino-dominated one in that both $C_{\tilde{\chi}_1^0 \tilde{\chi}_1^0 h}$ and $C_{\tilde{\chi}_1^0 \tilde{\chi}_1^0 Z}$ are proportional to $\lambda^2$ when the LHC-discovered scalar has the same properties as the SM Higgs boson~\cite{Baum:2017enm,Zhou:2021pit,Cao:2021ljw}, where $\lambda$ denotes the singlet-doublet Higgs Yukawa coupling in superpotential. Evidently, this type of DM can readily satisfy the constraints from the DM direct detection experiments when $\lambda$ is small. In the $Z_3$-NMSSM, the DM usually obtains the measured relic density either by annihilating into $t \bar{t}$ state through an off-shell $Z$-boson exchange or by co-annihilating with higgsino-dominated electroweakinos~\cite{Baum:2017enm,Zhou:2021pit}. The former case happens for $0.4 \lesssim \lambda \lesssim 0.7$. Although its prediction on the SI cross-section may be less than $10^{-47} {\rm cm^2}$ due to the cancellation of different contributions shown in Fig. 2 of Ref.~\cite{Zhou:2021pit}, the SD cross-section is always at the order of $10^{-41} {\rm cm^2}$, which contradicts to the recent LZ experimental limits. The latter case is tenable within a narrow parameter space characterized by $2 |\kappa|/\lambda \simeq 1$, $\lambda \lesssim 0.1$, and $\mu \lesssim 400~{\rm GeV}$, where $\kappa$ parameterizes the singlet Higgs field's self-coupling. Even without the demand to explain the muon g-2 anomaly, its Bayesian evidence is heavily suppressed, indicating that the case needs tuning of its parameters to survive experimental constraints.
In addition, it should be noted that the singlino-dominated dark matter may also achieve the measured density through the resonant annihilation into singlet-dominated CP-even or -odd Higgs boson~\cite{Belanger:2005kh,
 Cao:2009ad,Mahmoudi:2010xp,Cao:2011re,Cao:2013mqa,Han:2014nba,Cheung:2014lqa,Ellwanger:2014hia,Cao:2014efa,Cao:2015loa,Abdallah:2019znp,Wang:2020dtb,Barman:2020vzm,Ahmed:2022jlo}. This mechanism, however, becomes less favored by the improving DM direct detection experiments. The fundamental reason is that smaller and smaller $\lambda$  and $\kappa$ (note that $|\kappa| < \lambda/2$ in the $Z_3$-NMSSM to obtain the singlino-dominated DM), as required by the direct detection experiments, will reduce the DM's coupling strength with the scalars. As a result, $2 |m_{\tilde{\chi}_1^0}|$ must be closer to the scalar mass to obtain the measured density. This situation requires the fine-tuning quantity, defined in Eq.(19) of Ref.~\cite{Cao:2018rix} to get the measured density, to be larger than about 150. It will deteriorate if a smaller density is demanded to avoid the over-closure of the universe. So the mechanism contributes little to the Bayesian evidence in an elaborated scan of the model's parameter space with the MultiNest algorithm~\cite{Feroz:2008xx} and is, therefore, usually neglected. In the footnote 6 of this work, we will discuss more about this subject.

The deficit of the $Z_3$-NMSSM is amended in the general NMSSM (GNMSSM), where the relationship $2 |\kappa|/\lambda < 1$ in the $Z_3$-NMSSM does not hold~\cite{Cao:2021ljw}. Specifically, the singlet-dominated particles, such as the singlino-dominated DM and singlet-dominated Higgs bosons, can form a secluded DM sector, where $\kappa$ controls the couplings among these particles. In this case, the singlino-dominated DM achieves the measured relic density by annihilating into a pair of the singlet-dominated Higgs bosons~\cite{Cao:2021ljw}. Other distinct features of the GNMSSM include~\cite{Cao:2021tuh}
\begin{itemize}
\item The LHC restrictions on sparticle mass spectra are significantly relaxed since heavy sparticles tend to decay first into lighter sparticles other than the singlino-dominated DM. Consequently, the decay chain is lengthened, and the decay product becomes complicated.
\item The muon g-2 anomaly can be explained by light higgsinos, favored in natural electroweak symmetry breaking. This situation is in sharp contrast with the bino-dominated DM case.
\item The DM direct detection experiments prefer a small $\lambda$, which is beneficial to stabilize the vacuum of the scalar potential in the GNMSSM.
\end{itemize}

In the GNMSSM, the SM-like Higgs boson may be the lightest CP-even Higgs state or next-to-lightest CP-even Higgs state, which is dubbed as $h_1$-scenario and $h_2$-scenario, respectively, in literature. We showed in Ref.~\cite{Cao:2022htd} that the $h_2$-scenario is featured by $\tan \beta \lesssim 30$ and moderately light higgsinos, $\mu_{tot} \lesssim 500~{\rm GeV}$. If required to account for the muon g-2 anomaly, it entails some light sparticles which make the LHC constraints very tight. As a result, the scenario is hard to keep consistent with all experimental data. Therefore, we concentrate on the $h_1$-scenario in this study to survey the status of the singlino-dominated DM, given the great progress of the LHC experiments, the DM direct detection experiments, and the muon g-2 measurement in recent years. This work differs from the previous study of $\mu$-term extended NMSSM ($\mu$NMSSM) in following aspects~\cite{Cao:2021tuh,Cao:2022htd}. First, the GNMSSM comprises the $\mu$NMSSM and thus provides more flexible mechanisms to be compatible with experiments. For example, the mass for the singlet-dominated Higgs bosons in the GNMSSM can be treated as free parameters. Thus, the theory opens up more DM annihilation channels than the $\mu$NMSSM to affect the relic density. Second, some additional analyses of the LHC data, such as the latest ATLAS search for tri-lepton + $E_T^{\rm miss}$ signal~\cite{ATLAS:2021moa}, are implemented in this study. They provide more stringent constraints on the theory. Last, the impact of the recent LZ experiment on the GNMSSM is scrutinized, which can further strengthen the experimental restrictions on the GNMSSM.

This work is organized as follows. In Sec. \ref{theory-section}, we briefly introduce the basics of GNMSSM and the SUSY contribution to $a_\mu$. In Sec. \ref{numerical study}, we perform a sophisticated scan over the broad parameter space of the GNMSSM and investigate the status of the singlino-dominated DM in view of the experimental advancements. We also survey the impact of the recent LZ experiment on the GNMSSM. Lastly, we draw conclusions in Sec. \ref{conclusion-section}.

\section{\label{theory-section}Theoretical preliminaries of GNMSSM}

The superpotential of the GNMSSM is given by~\cite{Ellwanger:2009dp,Maniatis:2009re,Fayet:1974pd}
\begin{eqnarray}
W_{\rm GNMSSM} = W_{\rm Yukawa} + \lambda \hat{S}\hat{H_u} \cdot \hat{H_d} + \frac{\kappa}{3}\hat{S}^3 + \mu \hat{H_u} \cdot \hat{H_d} + \xi\hat{S} + \frac{1}{2} \mu^\prime \hat{S}^2,  \label{Superpotential}
\end{eqnarray}
where the Yukawa terms contained in $W_{\rm Yukawa}$ are the same as those of the MSSM, $\hat{H}_u=(\hat{H}_u^+,\hat{H}_u^0)^T$ and $\hat{H}_d=(\hat{H}_d^0,\hat{H}_d^-)^T$ represent $SU(2)_L$ doublet Higgs superfields, and $\lambda$, $\kappa$ are dimensionless coupling coefficients parameterizing the $Z_3$-invariant trilinear terms. Historically, the $Z_3$-symmetry violating terms, characterized by the bilinear mass parameters $\mu$, $\mu^\prime$ and the singlet tadpole parameter $\xi$, were introduced to solve the tadpole problem~\cite{Ellwanger:1983mg, Ellwanger:2009dp} and the cosmological domain-wall problem of the $Z_3$-NMSSM~\cite{Abel:1996cr, Kolda:1998rm, Panagiotakopoulos:1998yw}. The bilinear terms might stem from an underlying discrete R symmetry, $Z^R_4$ or $Z^R_8$, after SUSY breaking, and could be naturally at the electroweak scale~\cite{Abel:1996cr,Lee:2010gv,Lee:2011dya,Ross:2011xv,Ross:2012nr}. They can change significantly the properties of the Higgs bosons and the neutralinos in the $Z_3$-NMSSM  and lead to much richer phenomenology.  The
$\xi$-term can be eliminated by shifting the $\hat{S}$ field with a constant and redefining the $\mu$ parameter~\cite{Ross:2011xv}. So, without loss of generality, $\xi$ is set to be zero in this study.

\subsection{Higgs sector}

The soft-breaking terms for the Higgs fields take the following form~\cite{Ellwanger:2009dp,Maniatis:2009re}:
\begin{align}
-\mathcal{L}_{soft} = &\Bigg[\lambda A_{\lambda} S H_u \cdot H_d + \frac{1}{3} A_{\kappa} \kappa S^3+ m_3^2 H_u\cdot H_d + \frac{1}{2} {m_S^{\prime}}^2 S^2 + h.c.\Bigg] \nonumber \\
& + m^2_{H_u}|H_u|^2 + m^2_{H_d}|H_d|^2 + m^2_{S}|S|^2 ,
\end{align}
where $H_u$, $H_d$ and $S$ represent the scalar components of the Higgs superfields, and their squared masses, $m_{H_u}^2$, $m_{H_d}^2$ and $m_{S}^2$, can be fixed by solving the conditional equations to minimize the scalar potential and expressed in terms of the vevs of the Higgs fields, which are denoted as $\left\langle H_u^0 \right\rangle = v_u/\sqrt{2}$, $\left\langle H_d^0 \right\rangle = v_d/\sqrt{2}$ and $\left\langle S \right\rangle = v_s/\sqrt{2}$ with $v = \sqrt{v_u^2+v_d^2}\simeq 246~\mathrm{GeV}$. As usual, an effective $\mu$-parameter is defined by $\mu_{\rm eff} \equiv \lambda v_s/\sqrt{2}$. After these arrangements, the Higgs sector is described by ten free parameters: $\tan \beta$, the Yukawa couplings $\lambda$ and $\kappa$, $\mu_{eff}$, the soft-breaking trilinear coefficients $A_\lambda$ and $A_\kappa$, the bilinear mass parameters $\mu$ and $\mu^\prime$, and their soft-breaking parameters $m_3^2$ and $m_S^{\prime\ 2}$.

In revealing the characteristics of the Higgs physics, it is customary to work with the following field combinations: $H_{\rm SM} \equiv \sin\beta {\rm Re}(H_u^0) + \cos\beta {\rm Re} (H_d^0)$, $H_{\rm NSM} \equiv \cos\beta {\rm Re}(H_u^0) - \sin\beta {\rm Re}(H_d^0)$, and $A_{\rm NSM} \equiv \cos\beta {\rm Im}(H_u^0) - \sin\beta  {\rm Im}(H_d^0)$, where $H_{\rm SM}$ stands for the SM Higgs field, and $H_{\rm NSM}$ and $A_{\rm NSM}$ represent the beyond-SM doublet fields~\cite{Cao:2012fz}. The elements of $CP$-even Higgs boson mass matrix $\mathcal{M}_S^2$ in the bases $\left(H_{\rm NSM}, H_{\rm SM}, {\rm Re}[S]\right)$ are read as follows~\cite{Ellwanger:2009dp}:
\begin{eqnarray}\label{CP-even Hisggs Mass}
  {\cal M}^2_{S, 11}&=& \frac{2 \left [ \mu_{eff} (\lambda A_\lambda + \kappa \mu_{eff} + \lambda \mu^\prime ) + \lambda m_3^2 \right ] }{\lambda \sin 2 \beta} + \frac{1}{2} (2 m_Z^2- \lambda^2v^2)\sin^22\beta, \nonumber \\
  {\cal M}^2_{S, 12}&=&-\frac{1}{4}(2 m_Z^2-\lambda^2v^2)\sin4\beta, \nonumber \\
  {\cal M}^2_{S, 13}&=&-\frac{1}{\sqrt{2}} ( \lambda A_\lambda + 2 \kappa \mu_{eff} + \lambda \mu^\prime ) v \cos 2 \beta, \nonumber \\
  {\cal M}^2_{S, 22}&=&m_Z^2\cos^22\beta+ \frac{1}{2} \lambda^2v^2\sin^22\beta,\nonumber  \\
  {\cal M}^2_{S, 23}&=& \frac{v}{\sqrt{2}} \left[2 \lambda (\mu_{eff} + \mu) - (\lambda A_\lambda + 2 \kappa \mu_{eff} + \lambda \mu^\prime ) \sin2\beta \right], \nonumber \\
  {\cal M}^2_{S, 33}&=& \frac{\lambda (A_\lambda + \mu^\prime) \sin 2 \beta}{4 \mu_{eff}} \lambda v^2   + \frac{\mu_{eff}}{\lambda} (\kappa A_\kappa +  \frac{4 \kappa^2 \mu_{eff}}{\lambda} + 3 \kappa \mu^\prime ) - \frac{\mu}{2 \mu_{eff}} \lambda^2 v^2, \quad \label{Mass-CP-even-Higgs}
\end{eqnarray}
and those for $CP$-odd Higgs fields in the bases $\left( A_{\rm NSM}, {\rm Im}(S)\right)$ are given by Equations~(\ref{Mass-CP-odd-Higgs}):
\begin{eqnarray}
{\cal M}^2_{P,11}&=& \frac{2 \left [ \mu_{eff} (\lambda A_\lambda + \kappa \mu_{eff} + \lambda \mu^\prime ) + \lambda m_3^2 \right ] }{\lambda \sin 2 \beta}, \nonumber  \\
{\cal M}^2_{P,22}&=& \frac{(\lambda A_\lambda + 4 \kappa \mu_{eff} + \lambda \mu^\prime ) \sin 2 \beta }{4 \mu_{eff}} \lambda v^2  - \frac{\kappa \mu_{eff}}{\lambda} (3 A_\kappa + \mu^\prime) - \frac{\mu}{2 \mu_{eff}} \lambda^2 v^2 - 2 m_S^{\prime\ 2}, \nonumber  \\
{\cal M}^2_{P,12}&=& \frac{v}{\sqrt{2}} ( \lambda A_\lambda - 2 \kappa \mu_{eff} - \lambda \mu^\prime ). \label{Mass-CP-odd-Higgs}
\end{eqnarray}
The mass eigenstates $h_i=\{h, H, h_s\}$ and $a_i=\{A_H, A_s\}$ are obtained by unitary rotations $V$ and $V_P$ to diagonalize ${\cal{M}}_S^2$ and ${\cal{M}}_P^2$, respectively, leading to the following decompositions:
\begin{eqnarray} \label{Mass-eigenstates}
h_i & = & V_{h_i}^{\rm NSM} H_{\rm NSM}+V_{h_i}^{\rm SM} H_{\rm SM}+V_{h_i}^{\rm S} Re[S], \quad \quad a_i =  V_{P, a_i}^{\rm NSM} A_{\rm NSM}+ V_{P, a_i}^{\rm S} Im [S].
\end{eqnarray}
Among these states, $h$ denotes the SM-like scalar discovered at the LHC, $H$ and $A_H$ represent the heavy doublet-dominated states, and $h_s$ and $A_s$ are the singlet-dominated states. For convenience, these states are also labelled in ascending mass orders, i.e. $m_{h_1} < m_{h_2} < m_{h_3}$, and $m_{A_1} < m_{A_2}$. Thus, $h \equiv h_1$ and $m_{h_s} > m_h$ in the $h_1$-scenario. The mass of the charged Higgs state $H^\pm = \cos \beta H_u^\pm + \sin \beta H_d^\pm$ is given in Equation~(\ref{charged-Higgs}):
\begin{eqnarray}  \label{charged-Higgs}
m^2_{H^{\pm}} &=& m_A^2 +  m^2_W -\lambda^2 v^2,
\end{eqnarray}
where $m_A^2 \equiv 2 \left [ \mu_{eff} (\lambda A_\lambda + \kappa \mu_{eff} + \lambda \mu^\prime ) + \lambda m_3^2 \right ]/(\lambda \sin 2 \beta)$ represents the squared mass of the $A_{\rm NSM}$ field. So far, extra Higgs bosons, $H$, $A_H$, $h_s$, $A_s$ and $H^\pm$, have been  searched for intensively at the LHC (see, e.g., Ref.~\cite{ATLAS:2020zms,ATLAS:2021upq}), and model-independent upper limits on their production rate are acquired.

In the limit of $\lambda \to 0$, $m_A^2 \simeq 2 m_3^2/\sin 2 \beta$, and the mass matrix elements are approximated by\footnote{Given $\mu_{eff} \equiv \lambda v_s/\sqrt{2}$ and $\mu_{eff} \lesssim 30 {\rm GeV}$ indicated by numerical results of this study (see discussions below), we also neglect terms proportional to $\mu_{eff}$ in the mass matrices. It agrees with the approximation of $2 \kappa \mu_{eff} \simeq \lambda m_{\tilde{\chi}_1^0} - \lambda \mu^\prime$, indicated by the first expression in Eq.~(\ref{Approximation-neutralinos}). }
\begin{eqnarray}
  {\cal M}^2_{S, 11} & \simeq & m_A^2 + m_Z^2 \sin^22\beta, \quad \quad {\cal M}^2_{S, 12} \simeq -\frac{1}{2} m_Z^2 \sin4\beta, \quad \quad {\cal M}^2_{S, 13} \simeq 0,  \nonumber \\
  {\cal M}^2_{S, 22} &\simeq & m_Z^2\cos^22\beta,\quad \quad  {\cal M}^2_{S, 23} \simeq 0, \quad \quad {\cal M}^2_{S, 33} \simeq  \frac{\mu_{eff}}{\lambda} (\kappa A_\kappa +  \frac{4 \kappa^2 \mu_{eff}}{\lambda} + 3 \kappa \mu^\prime ), \nonumber \\
 {\cal M}^2_{P,11}& = & m_A^2, \quad \quad {\cal M}^2_{P,22} \simeq  - \frac{\kappa \mu_{eff}}{\lambda} (3 A_\kappa + \mu^\prime) - 2 m_S^{\prime\ 2}, \quad \quad {\cal M}^2_{P,12} \simeq 0.
\end{eqnarray}
These formulae reflect the following facts~\cite{Cao:2022htd}:
\begin{itemize}
\item Parameters $A_\lambda$ and $m_3$ determine the masses of the heavy doublet-dominated scalars. Given a small $\lambda$ preferred by the LZ restriction on the singlino-dominated DM scenario, they affect little the other Higgs bosons' properties.
\item Parameters $A_\kappa$ and $m_S^{\prime}$ appear only in ${\cal M}^2_{S, 33}$ and ${\cal M}^2_{P,22}$, which implies that $m_{h_s}$ and $m_{A_s}$ can vary freely by adjusting $A_\kappa$ and $m_S^{\prime}$. This situation is different from that of the $Z_3$-NMSSM, where $\mu_{tot} \equiv \mu_{eff}$, $\mu^\prime = 0$, and $m_S^{\prime}=0$, and consequently the masses of singlet fields are correlated~\cite{Cao:2018rix}. In this study, we are particularly interested in the mass hierarchy $m_{h_s},m_{A_s} \ll m_A$ since the singlet-dominated scalars play crucial roles in the DM annihilation.
\end{itemize}

Given that too many parameters are involved in the Higgs sector, we study the $h_1$-scenario by assuming the charged Higgs bosons to be massive through setting $A_\lambda = 500~{\rm GeV}$ and $m_3 = 250~{\rm GeV}$. Specifically, it is found that $m_{H^\pm}$ ranges from about $1050~{\rm GeV}$ to about $5000~{\rm GeV}$ in this study, which is consistent with the constraints from the LHC search for $H^\pm$~\cite{ATLAS:2021upq}.

\subsection{Neutralino sector}

The neutralino sector of the GNMSSM comprises bino field $\tilde{B}$, wino field $\tilde{W}$, higgsino fields $\tilde{H}_d^0$ and $\tilde{H}_u^0$, and singlino field $\tilde{S}$. Its mass matrix in the bases $(-i \tilde{B}, -i \tilde{W},\tilde{H}_d^0,\tilde{H}_u^0,\tilde{S})$ takes the following form ~\cite{Ellwanger:2009dp}, as shown in Equation~(\ref{eq:mmn}):
\begin{equation}
    M_{\tilde{\chi}^0} = \left(
    \begin{array}{ccccc}
    M_1 & 0 & -m_Z \sin \theta_W \cos \beta & m_Z \sin \theta_W \sin \beta & 0 \\
      & M_2 & m_Z \cos \theta_W \cos \beta & - m_Z \cos \theta_W \sin \beta &0 \\
    & & 0 & -\mu_{tot} & - \frac{1}{\sqrt{2}} \lambda v \sin \beta \\
    & & & 0 & -\frac{1}{\sqrt{2}} \lambda v \cos \beta \\
    & & & & \frac{2 \kappa}{\lambda} \mu_{\rm eff} + \mu^\prime
    \end{array}
    \right), \label{eq:mmn}
\end{equation}
where $M_1$ and $M_2$ denote gaugino soft-breaking masses, and $\mu_{tot}\equiv \mu_{\rm eff} + \mu$ represents the higgsino mass. With a rotation matrix $N$ to diagonalize this mass matrix, neutralino mass eigenstates are expressed by Equation~(\ref{Mass-eigenstate-neutralino}):
\begin{eqnarray}  \label{Mass-eigenstate-neutralino}
\tilde{\chi}_i^0 = N_{i1} \psi^0_1 +   N_{i2} \psi^0_2 +   N_{i3} \psi^0_3 +   N_{i4} \psi^0_4 +   N_{i5} \psi^0_5.
\end{eqnarray}
where $\tilde{\chi}_i^0\,(i=1,2,3,4,5)$ are labeled in an ascending mass order, $N_{i3}$ and $N_{i4}$ characterize the $\tilde{H}_d^0$ and $\tilde{H}_u^0$ components in $\tilde{\chi}_i^0$, respectively, and $N_{i5}$ represents the singlino component.

Assuming $|m_{\tilde{\chi}_1^0}^2 - \mu_{tot}^2 | \gg \lambda^2 v^2$ and very massive gauginos\footnote{Note that these assumptions are unnecessary in the $\lambda \to 0$ limit encountered in this work, where the singlino field decouples from the rest fields in the neutralino sector, and $m_{\tilde{\chi}_1^0} \simeq \sqrt{2} \kappa v_s + \mu^\prime $. }, the mass of the singlino-dominated $\tilde{\chi}_1^0$ and its field compositions are approximated by~\cite{Cheung:2014lqa,Badziak:2015exr,Badziak_2017}
\begin{eqnarray} \label{Approximation-neutralinos}
m_{\tilde{\chi}_1^0} & \simeq & \frac{2 \kappa}{\lambda} \mu_{eff} + \mu^\prime + \frac{1}{2} \frac{\lambda^2 v^2 ( m_{\tilde{\chi}_1^0} - \mu_{tot} \sin 2 \beta )}{m_{\tilde{\chi}_1^0}^2 - \mu_{tot}^2}, \quad N_{11} \sim 0, \quad N_{12} \sim 0, \label{Neutralino-Mixing}  \\
\frac{N_{13}}{N_{15}} &= & \frac{\lambda v}{\sqrt{2} \mu_{\rm tot}} \frac{(m_{\tilde{\chi}_1^0}/\mu_{\rm tot})\sin\beta-\cos\beta} {1-\left(m_{\tilde{\chi}_1^0}/\mu_{\rm tot}\right)^2}, \quad \quad  \frac{N_{14}}{N_{15}} =  \frac{\lambda v}{\sqrt{2} \mu_{\rm tot}} \frac{(m_{\tilde{\chi}_1^0}/\mu_{\rm tot})\cos\beta-\sin\beta} {1-\left(m_{\tilde{\chi}_1^0}/\mu_{\rm tot}\right)^2}, \nonumber \\
N_{15}^2 & \simeq & \left(1+ \frac{N^2_{13}}{N^2_{15}}+\frac{N^2_{14}}{N^2_{15}}\right)^{-1} \nonumber \\
&= & \frac{\left[1-(m_{\tilde{\chi}_1^0}/\mu_{\rm tot})^2\right]^2}{\left[(m_{\tilde{\chi}_1^0}/\mu_{\rm tot})^2
-2(m_{\tilde{\chi}_1^0}/\mu_{\rm tot})\sin2\beta+1 \right]\left(\frac{\lambda v}{\sqrt{2}\mu_{\rm tot}}\right)^2
+\left[1-(m_{\tilde{\chi}_1^0}/\mu_{\rm tot})^2\right]^2}. \nonumber
\end{eqnarray}
These analytic expressions show the following properties of the DM:
\begin{itemize}
\item $m_{\tilde{\chi}_1^0}$ is determined by parameters $\lambda$, $\kappa$, $\tan \beta$, $\mu_{eff}$, $\mu_{tot}$, and $\mu^\prime$. Even when $\lambda$, $\kappa$, $\tan \beta$, $\mu_{eff}$, and $\mu_{tot}$ are fixed, it can still vary freely by adjusting $\mu^\prime$. In addition, $\lambda$ and $\kappa$ in the GNMSSM are independent parameters in predicting $|m_{\tilde{\chi}_1^0}| < |\mu_{tot}|$. These situations are different from those of the $Z_3$-NMSSM, where $\mu^\prime \equiv 0$ and $\mu_{tot} \equiv \mu_{eff}$, and consequently $|\kappa|$ must be less than $\lambda/2$ to predict the singlino-dominated neutralino as the LSP~\cite{Ellwanger:2009dp}.
\item The field compositions in $\tilde{\chi}_1^0$ depend only on $\tan \beta$, $\lambda$, $\mu_{tot}$, and $m_{\tilde{\chi}_1^0}$. Therefore,  it is convenient to take the four parameters and $\kappa$ as theoretical inputs in studying the $\tilde{\chi}_1^0$'s properties, where $\kappa$ determines the interactions among the singlet-dominated particles. This feature contrasts with that of the $Z_3$-NMSSM, which needs only four input parameters, namely $\tan \beta$, $\lambda$, $\mu_{tot}$, and any of $m_{\tilde{\chi}_1^0}$ or $\kappa$, to describe the properties of $\tilde{\chi}_1^0$~\cite{Zhou:2021pit}.
\item The singlet-dominated particles may form a secluded DM sector~\cite{Pospelov:2007mp}, where the DM achieves the measured density by the process $\tilde{\chi}_1^0 \tilde{\chi}_1^0 \to h_s A_s, h_s h_s, A_s A_s$ or the $h_s/A_s$-mediated resonant annihilation into SM particles. The former annihilation proceeds mainly through the $t$-channel exchange of $\tilde{\chi}_1^0$ and/or the $s$-channel exchange of $h_s/A_s$, and thus, its cross-section is controlled by the parameters $\kappa$ and $A_\kappa$. The rate of the latter annihilation, however, also strongly depends on the splitting between $2|m_{\tilde{\chi}_1^0}|$ and $m_{h_s/A_s}$.
    Since this sector communicates with the SM sector only through the weak singlet-doublet Higgs mixing, the DM-nucleon scattering is naturally suppressed when $\lambda$ is small.
\end{itemize}

 Owing to these salient features, the GNMSSM predicts a broad parameter space consistent with current DM experimental results.

\subsection{\label{DMRD}Muon g-2 }

The SUSY source of the muon g-2, $a^{\rm SUSY}_{\mu}$, mainly includes loops mediated by a smuon and a neutralino and those containing a muon-type sneutrino and a chargino~\cite{Moroi:1995yh,Domingo:2008bb,Hollik:1997vb,Martin:2001st}. The full one-loop contributions to $a^{\rm SUSY}_{\mu}$ in the NMSSM were obtained in Ref.~\cite{Domingo:2008bb} and are not presented here for brevity. As an alternative, we provide the expression of $a_\mu^{\rm SUSY}$ in the mass insertion approximation~\cite{Moroi:1995yh} to reveal its key features. Specifically, at the lowest order of the approximation, the contributions to $a_\mu^{\rm SUSY}$ are divided into four types: "WHL", "BHL", "BHR", and "BLR", where $W$, $B$, $H$, $L$, and $R$ represent wino, bino, higgsino, and left-handed and right-handed smuon fields, respectively. They arise from the Feynman diagrams involving $\tilde{W}-\tilde{H}_d$, $\tilde{B}-\tilde{H}_d^0$, $\tilde{B}-\tilde{H}_d^0$, and $\tilde{\mu}_L-\tilde{\mu}_R$ transitions, respectively, and take the following form~\cite{Athron:2015rva, Moroi:1995yh,Endo:2021zal}:
\begin{eqnarray}
a_{\mu, \rm WHL}^{\rm SUSY}
    &=&\frac{\alpha_2}{8 \pi} \frac{m_{\mu}^2 M_2 \mu_{tot} \tan \beta}{m_{\tilde{\nu}_\mu}^4} \left \{ 2 f_C\left(\frac{M_2^2}{m_{\tilde{\nu}_{\mu}}^2}, \frac{\mu_{tot}^2}{m_{\tilde{\nu}_{\mu}}^2} \right) - \frac{m_{\tilde{\nu}_\mu}^4}{\tilde{m}_{\tilde{\mu}_L}^4} f_N\left(\frac{M_2^2}{\tilde{m}_{\tilde{\mu}_L}^2}, \frac{\mu_{tot}^2}{\tilde{m}_{\tilde{\mu}_L}^2} \right) \right \}\,, \quad \quad
    \label{eq:WHL} \\
a_{\mu, \rm BHL}^{\rm SUSY}
  &=& \frac{\alpha_Y}{8 \pi} \frac{m_\mu^2 M_1 \mu_{tot}  \tan \beta}{\tilde{m}_{\tilde{\mu}_L}^4} f_N\left(\frac{M_1^2}{\tilde{m}_{\tilde{\mu}_L}^2}, \frac{\mu_{tot}^2}{\tilde{m}_{\tilde{\mu}_L}^2} \right)\,,
    \label{eq:BHL} \\
a_{\mu, \rm BHR}^{\rm SUSY}
  &=& - \frac{\alpha_Y}{4\pi} \frac{m_{\mu}^2 M_1 \mu_{tot} \tan \beta}{\tilde{m}_{\tilde{\mu}_R}^4} f_N\left(\frac{M_1^2}{\tilde{m}_{\tilde{\mu}_R}^2}, \frac{\mu_{tot}^2}{\tilde{m}_{\tilde{\mu}_R}^2} \right)\,,
    \label{eq:BHR} \\
a_{\mu \rm BLR}^{\rm SUSY}
  &=& \frac{\alpha_Y}{4\pi} \frac{m_{\mu}^2  M_1 \mu_{tot} \tan \beta}{M_1^4}
    f_N\left(\frac{\tilde{m}_{\tilde{\mu}_L}^2}{M_1^2}, \frac{\tilde{m}_{\tilde{\mu}_R}^2}{M_1^2} \right)\,,
    \label{eq:BLR}
\end{eqnarray}
where $\tilde{m}_{\tilde{\mu}_L}$ and $\tilde{m}_{\tilde{\mu}_R}$ are soft-breaking masses for left-handed and right-handed smuon fields, respectively.  The loop functions are given by
\begin{eqnarray}
    \label{eq:loop-aprox}
    f_C(x,y)
    &=&  \frac{5-3(x+y)+xy}{(x-1)^2(y-1)^2} - \frac{2\ln x}{(x-y)(x-1)^3}+\frac{2\ln y}{(x-y)(y-1)^3} \,,
      \\
    f_N(x,y)
    &=&
      \frac{-3+x+y+xy}{(x-1)^2(y-1)^2} + \frac{2x\ln x}{(x-y)(x-1)^3}-\frac{2y\ln y}{(x-y)(y-1)^3} \,,
\end{eqnarray}
and they satisfy $f_C(1,1) = 1/2$ and $f_N(1,1) = 1/6$.

These approximations reveal the following facts:
\begin{itemize}
\item If all SUSY parameters involved in $a_\mu^{\rm SUSY}$ take a common value $M_{\rm SUSY}$, $a_\mu^{\rm SUSY}$ is proportional to $m_\mu^2 \tan \beta/M_{\rm SUSY}^2$, indicating that  the muon g-2 anomaly prefers a large $\tan \beta$ and a moderately low SUSY scale.
\item It was verified that the "WHL" contribution is usually much larger than the other contributions if $\tilde{\mu}_L$ is not significantly heavier than $\tilde{\mu}_R$~\cite{Cao:2021tuh}.
\item In principle, the singlino field $\tilde{S}$ enters the insertions. However, since both the $\tilde{W}-\tilde{S}$ and $\tilde{B}^0-\tilde{S}$ transitions and the $\bar{\mu} \tilde{S} \tilde{\mu}_{L,R}$ couplings vanish~\cite{Ellwanger:2009dp}, it only appears in the "WHL", "BHL" and "BHR" loops by two more insertions at the lowest order, which correspond to the $\tilde{H}_d^0-\tilde{S}$ and $\tilde{S}-\tilde{H}_d^0$ transitions in the neutralino mass matrix in Eq.~(\ref{eq:mmn}). Since the DM physics prefers a small $\lambda$ and the LHC search for SUSY places a lower bound of about $140~{\rm GeV}$ for the singlino mass (see the following discussions), the singlino-induced contributions are never significant~\cite{Cao:2021tuh}. Specifically, we survey the properties of lots of parameter points acquired in this study.
    For each point, we compare different $a_\mu^{\rm SUSY}$, obtained by its exact formula, by that but fixing $\lambda$ at a small value of $0.005$, and by the mass insertion method that entirely neglects the dependence of $a_\mu^{\rm SUSY}$ on $\lambda$, respectively. We find that the differences between these predictions are within $5\%$ for all the considered points.

\item Although the GNMSSM prediction of $a_\mu^{\rm SUSY}$ is approximately the same as that of the MSSM, except that the $\mu$
parameter in the MSSM is replaced by $\mu_{tot}$, the two models predict different DM physics and different sparticle signals at the LHC. As a result, they are
subject to different theoretical and experimental constraints.
\end{itemize}

In this study, we do not consider two-loop (2L) contributions to $a_\mu$, which include 2L corrections to SM one-loop diagrams and those to SUSY one-loop diagrams~\cite{Stockinger:2006zn}. Before the advent of the LHC, these contributions were estimated to be significant in the parameter space characterized by moderately light $H/A_H$ and sparticles~\cite{Stockinger:2006zn}. We utilize code GM2Calc~\cite{Athron:2015rva} to study them by selecting several MSSM benchmark points from a related study (see footnote 2 of this work)\footnote{Given that the code GM2Calc works only in the framework of MSSM~\cite{Athron:2015rva} and that GNMSSM predicts approximately the same $a_\mu^{\rm SUSY}$ as MSSM, results of the GM2Calc about the 2L corrections can be applied to GNMSSM. }. These points predict different mass orders for the $W$, $B$, $H$, $L$, and $R$ fields, but all correspond to  $\tan \beta > 50$ and $a_\mu^{1L} \simeq 2.5 \times 10^{-9}$, with $a_\mu^{1L}$ denoting the one-loop result of $a_\mu^{\rm SUSY}$. We find $a_\mu^{\rm 2L, photonic} \simeq - 2.0 \times 10^{-10}$, $a_\mu^{\rm 2L, f\tilde{f}} \simeq (6.0 \sim 9.0 ) \times 10^{-11}$, and $a_\mu^{\rm 2L (a)} \sim 10^{-12}$, where $a_\mu^{\rm 2L, photonic}$, $a_\mu^{\rm 2L, f\tilde{f}}$, and $a_\mu^{\rm 2L (a)}$ denote photonic 2L corrections, fermion/sfermion corrections, and corrections from the photonic Barr-Zee diagrams involving pure SUSY loops of charginos, neutralinos, and sfermions, respectively~\cite{Athron:2015rva}. The total 2L effect is about $-5 \%$ of $a_\mu^{1L}$. In addition, we note that the code GM2Calc does not include  contributions from the Barr-Zee diagrams containing top/bottom quark loops, which were denoted by $a_\mu^{\rm SUSY, ferm, 2L}$ in Ref.~\cite{Stockinger:2006zn}. We calculate these contributions by the formulae in Ref.~\cite{Cheung:2001hz}. We conclude that for $\tan \beta = 60$ and $m_A = 1~{\rm TeV}$, the corrections are $-1.40 \times 10^{-11}$ for the $f\gamma H$ diagrams and $1.68 \times 10^{-11}$ for the $f\gamma A_H$ diagrams. These results reflect that the 2L contributions missed in this study affect little our conclusions.

\begin{table}[]
	\caption{Experimental analyses of the electroweakino production processes considered in this study, which are categorized by the topologies of the SUSY signals.}
	\label{Table1}
	\vspace{0.2cm}
	\resizebox{0.98\textwidth}{!}{
		\begin{tabular}{llll}
			\hline\hline
			\texttt{Scenario} & \texttt{Final State} &\multicolumn{1}{c}{\texttt{Name}}\\\hline
			\multirow{6}{*}{$\tilde{\chi}_{2}^0\tilde{\chi}_1^{\pm}\rightarrow WZ\tilde{\chi}_1^0\tilde{\chi}_1^0$}&\multirow{6}{*}{$n\ell (n\geq2) + nj(n\geq0) + \text{E}_\text{T}^{\text{miss}}$}&\texttt{CMS-SUS-20-001($137fb^{-1}$)}~\cite{CMS:2020bfa}\\&&\texttt{ATLAS-2106-01676($139fb^{-1}$)}~\cite{ATLAS:2021moa}\\&&\texttt{CMS-SUS-17-004($35.9fb^{-1}$)}~\cite{CMS:2018szt}\\&&\texttt{CMS-SUS-16-039($35.9fb^{-1}$)}~\cite{CMS:2017moi}\\&&\texttt{ATLAS-1803-02762($36.1fb^{-1}$)}~\cite{ATLAS:2018ojr}\\&&\texttt{ATLAS-1806-02293($36.1fb^{-1}$)}~\cite{ATLAS:2018eui}\\\\
			\multirow{2}{*}{$\tilde{\chi}_2^0\tilde{\chi}_1^{\pm}\rightarrow \ell\tilde{\nu}\ell\tilde{\ell}$}&\multirow{2}{*}{$n\ell (n=3) + \text{E}_\text{T}^{\text{miss}}$}&\texttt{CMS-SUS-16-039($35.9fb^{-1}$)}~\cite{CMS:2017moi}\\&&\texttt{ATLAS-1803-02762($36.1fb^{-1}$)}~\cite{ATLAS:2018ojr}\\\\
			$\tilde{\chi}_2^0\tilde{\chi}_1^{\pm}\rightarrow \tilde{\tau}\nu\ell\tilde{\ell}$&$2\ell + 1\tau + \text{E}_\text{T}^{\text{miss}}$&\texttt{CMS-SUS-16-039($35.9fb^{-1}$)}~\cite{CMS:2017moi}\\\\
			$\tilde{\chi}_2^0\tilde{\chi}_1^{\pm}\rightarrow \tilde{\tau}\nu\tilde{\tau}\tau$&$3\tau + \text{E}_\text{T}^{\text{miss}}$&\texttt{CMS-SUS-16-039($35.9fb^{-1}$)}~\cite{CMS:2017moi}\\\\
			\multirow{6}{*}{$\tilde{\chi}_{2}^0\tilde{\chi}_1^{\pm}\rightarrow Wh\tilde{\chi}_1^0\tilde{\chi}_1^0$}&\multirow{6}{*}{$n\ell(n\geq1) + nb(n\geq0) + nj(n\geq0) + \text{E}_\text{T}^{\text{miss}}$}&\texttt{ATLAS-1909-09226($139fb^{-1}$)}~\cite{ATLAS:2020pgy}\\&&\texttt{CMS-SUS-17-004($35.9fb^{-1}$)}~\cite{CMS:2018szt}\\&&\texttt{CMS-SUS-16-039($35.9fb^{-1}$)}~\cite{CMS:2017moi}\\
			&&\texttt{ATLAS-1812-09432($36.1fb^{-1}$)}\cite{ATLAS:2018qmw}\\&&\texttt{CMS-SUS-16-034($35.9fb^{-1}$)}\cite{CMS:2017kxn}\\&&\texttt{CMS-SUS-16-045($35.9fb^{-1}$)}~\cite{CMS:2017bki}\\\\
			\multirow{2}{*}{$\tilde{\chi}_1^{\mp}\tilde{\chi}_1^{\pm}\rightarrow WW\tilde{\chi}_1^0 \tilde{\chi}_1^0$}&\multirow{2}{*}{$2\ell + \text{E}_\text{T}^{\text{miss}}$}&\texttt{ATLAS-1908-08215($139fb^{-1}$)}~\cite{ATLAS:2019lff}\\&&\texttt{CMS-SUS-17-010($35.9fb^{-1}$)}~\cite{CMS:2018xqw}\\\\
			\multirow{2}{*}{$\tilde{\chi}_1^{\mp}\tilde{\chi}_1^{\pm}\rightarrow 2\tilde{\ell}\nu(\tilde{\nu}\ell)$}&\multirow{2}{*}{$2\ell + \text{E}_\text{T}^{\text{miss}}$}&\texttt{ATLAS-1908-08215($139fb^{-1}$)}~\cite{ATLAS:2019lff}\\&&\texttt{CMS-SUS-17-010($35.9fb^{-1}$)}~\cite{CMS:2018xqw}\\\\
			$\tilde{\chi}_2^{0}\tilde{\chi}_1^{\mp}\rightarrow h/ZW\tilde{\chi}_1^0\tilde{\chi}_1^0,\tilde{\chi}_1^0\rightarrow \gamma/Z\tilde{G}$&\multirow{2}{*}{$2\gamma + n\ell(n\geq0) + nb(n\geq0) + nj(n\geq0) + \text{E}_\text{T}^{\text{miss}}$}&\multirow{2}{*}{\texttt{ATLAS-1802-03158($36.1fb^{-1}$)}~\cite{ATLAS:2018nud}}\\$\tilde{\chi}_1^{\pm}\tilde{\chi}_1^{\mp}\rightarrow WW\tilde{\chi}_1^0\tilde{\chi}_1^0,\tilde{\chi}_1^0\rightarrow \gamma/Z\tilde{G}$&&\\\\
			$\tilde{\chi}_2^{0}\tilde{\chi}_1^{\pm}\rightarrow ZW\tilde{\chi}_1^0\tilde{\chi}_1^0,\tilde{\chi}_1^0\rightarrow h/Z\tilde{G}$&\multirow{4}{*}{$n\ell(n\geq4) + \text{E}_\text{T}^{\text{miss}}$}&\multirow{4}{*}{\texttt{ATLAS-2103-11684($139fb^{-1}$)}~\cite{ATLAS:2021yyr}}\\$\tilde{\chi}_1^{\pm}\tilde{\chi}_1^{\mp}\rightarrow WW\tilde{\chi}_1^0\tilde{\chi}_1^0,\tilde{\chi}_1^0\rightarrow h/Z\tilde{G}$&&\\$\tilde{\chi}_2^{0}\tilde{\chi}_1^{0}\rightarrow Z\tilde{\chi}_1^0\tilde{\chi}_1^0,\tilde{\chi}_1^0\rightarrow h/Z\tilde{G}$&&\\$\tilde{\chi}_1^{\mp}\tilde{\chi}_1^{0}\rightarrow W\tilde{\chi}_1^0\tilde{\chi}_1^0,\tilde{\chi}_1^0\rightarrow h/Z\tilde{G}$&&\\\\
			\multirow{3}{*}{$\tilde{\chi}_{i}^{0,\pm}\tilde{\chi}_{j}^{0,\mp}\rightarrow \tilde{\chi}_1^0\tilde{\chi}_1^0+\chi_{soft}\rightarrow ZZ/H\tilde{G}\tilde{G}$}&\multirow{3}{*}{$n\ell(n\geq2) + nb(n\geq0) + nj(n\geq0) + \text{E}_\text{T}^{\text{miss}}$}&\texttt{CMS-SUS-16-039($35.9fb^{-1}$)}~\cite{CMS:2017moi}\\&&\texttt{CMS-SUS-17-004($35.9fb^{-1}$)}~\cite{CMS:2018szt}\\&&\texttt{CMS-SUS-20-001($137fb^{-1}$)}~\cite{CMS:2020bfa}\\\\
			\multirow{2}{*}{$\tilde{\chi}_{i}^{0,\pm}\tilde{\chi}_{j}^{0,\mp}\rightarrow \tilde{\chi}_1^0\tilde{\chi}_1^0+\chi_{soft}\rightarrow HH\tilde{G}\tilde{G}$}&\multirow{2}{*}{$n\ell(n\geq2) + nb(n\geq0) + nj(n\geq0) + \text{E}_\text{T}^{\text{miss}}$}&\texttt{CMS-SUS-16-039($35.9fb^{-1}$)}~\cite{CMS:2017moi}\\&&\texttt{CMS-SUS-17-004($35.9fb^{-1}$)}~\cite{CMS:2018szt}\\\\
			$\tilde{\chi}_{2}^{0}\tilde{\chi}_{1}^{\pm}\rightarrow W^{*}Z^{*}\tilde{\chi}_1^0\tilde{\chi}_1^0$&$3\ell + \text{E}_\text{T}^{\text{miss}}$&\texttt{ATLAS-2106-01676($139fb^{-1}$)}~\cite{ATLAS:2021moa}\\\\
			\multirow{3}{*}{$\tilde{\chi}_{2}^{0}\tilde{\chi}_{1}^{\pm}\rightarrow Z^{*}W^{*}\tilde{\chi}_1^0\tilde{\chi}_1^0$}&\multirow{2}{*}{$2\ell + nj(n\geq0) + \text{E}_\text{T}^{\text{miss}}$}&\texttt{ATLAS-1911-12606($139fb^{-1}$)}~\cite{ATLAS:2019lng}\\&&\texttt{ATLAS-1712-08119($36.1fb^{-1}$)}~\cite{ATLAS:2017vat}\\&&\texttt{CMS-SUS-16-048($35.9fb^{-1}$)}~\cite{CMS:2018kag}\\\\
			\multirow{3}{*}{$\tilde{\chi}_{2}^{0}\tilde{\chi}_{1}^{\pm}+\tilde{\chi}_{1}^{\pm}\tilde{\chi}_{1}^{\mp}+\tilde{\chi}_{1}^{\pm}\tilde{\chi}_{1}^{0}$}&\multirow{3}{*}{$2\ell + nj(n\geq0) + \text{E}_\text{T}^{\text{miss}}$}&\texttt{ATLAS-1911-12606($139fb^{-1}$)}~\cite{ATLAS:2019lng}\\&&\texttt{ATLAS-1712-08119($36.1fb^{-1}$)}~\cite{ATLAS:2017vat}\\&&\texttt{CMS-SUS-16-048($35.9fb^{-1}$)}~\cite{CMS:2018kag}\\\hline
					
	\end{tabular}} % }
\end{table}

\begin{table}[]
	\caption{Same as Table~\ref{Table1}, but for the slepton production processes.}
	\label{Table2}
  \centering
	\vspace{0.2cm}
	\resizebox{0.7\textwidth}{!}{
		\begin{tabular}{llll}
			\hline\hline
			\texttt{Scenario} & \texttt{Final State} &\multicolumn{1}{c}{\texttt{Name}}\\\hline
\multirow{6}{*}{$\tilde{\ell}\tilde{\ell}\rightarrow \ell\ell\tilde{\chi}_1^0\tilde{\chi}_1^0$}&\multirow{6}{*}{$2\ell + \text{E}_\text{T}^{\text{miss}}$}&\multirow{1}{*}{\texttt{ATLAS-1911-12606($139fb^{-1}$)}~\cite{ATLAS:2019lng}}\\&&\multirow{1}{*}{\texttt{ATLAS-1712-08119($36.1fb^{-1}$)}~\cite{ATLAS:2017vat}}\\&&\multirow{1}{*}{\texttt{ATLAS-1908-08215($139fb^{-1}$)}~\cite{ATLAS:2019lff}}\\&&\multirow{1}{*}{\texttt{CMS-SUS-20-001($137fb^{-1}$)}~\cite{CMS:2020bfa}}\\&&\multirow{1}{*}{\texttt{ATLAS-1803-02762($36.1fb^{-1}$)}~\cite{ATLAS:2018ojr}}\\&&\multirow{1}{*}{\texttt{CMS-SUS-17-009($35.9fb^{-1}$)}~\cite{CMS:2018eqb}}\\\hline

\end{tabular}} % }
\end{table}

\vspace{-0.3cm}

\subsection{LHC search for SUSY}

Since some of the electroweakinos and sleptons involved in $a_\mu^{\rm SUSY}$ must be moderately light to account for the anomaly~\cite{Chakraborti:2020vjp}, they are copiously produced at the LHC and thus subjected to strong constraints from the SUSY searches at the LHC with $\sqrt{s}=13~\rm{TeV}$. These searches usually concentrate on the theories with the $R$-parity conservation~\cite{Fayet:1977yc,Farrar:1978xj}, where the LSP is undetected leading to missing energy in the final states.  Given the complexity of the production processes and decay modes, we scrutinize numerous signal topologies in implementing the restrictions.  In Tables~\ref{Table1} and~\ref{Table2}, we list the experimental analyses considered in this study. We find that the following reports are particularly critical:
\begin{itemize}
\item \texttt{CMS-SUS-16-039 and CMS-SUS-17-004~\cite{CMS:2017moi,CMS:2018szt}}: Search for electroweakino productions with two, three, or four leptons and missing transverse momentum ($\rm{E}_{\rm{T}}^{\rm{miss}}$) in the final states. One remarkable strategy of this analysis is that it includes all the possible final states and defines several categories by the number of leptons in the event, their flavors, and their charges to enhance the discovery potential. In the context of simplified models, the observed limit on wino-dominated $m_{\tilde{\chi}_1^{\pm}}$ in the chargino-neutralino production is about 650 GeV for the WZ topology, 480 GeV for the WH topology, and 535 GeV for the mixed topology.

\item \texttt{ATLAS-2106-01676~\cite{ATLAS:2021moa}}: Search for wino- or higgsino-dominated chargino-neutralino pair productions. This analysis investigates on-shell $WZ$, off-shell $WZ$, and $Wh$ categories in the decay chain and focuses on the final state containing exactly three leptons, possible ISR jets, and $\rm{E}_{\rm{T}}^{\rm{miss}}$.
    For the wino scenario in the simplified model, the exclusion bound of $m_{\tilde{\chi}_2^0}$ is about $640~\rm{GeV}$ for a massless $\tilde{\chi}_1^0$, and it is weakened
    as the mass difference between $\tilde{\chi}_2^0$ and $\tilde{\chi}_1^0$ diminishes. Specifically, $\tilde{\chi}_2^0$ should be heavier than about $500~\rm{GeV}$ for $m_{\tilde{\chi}_1^0} = 300~{\rm GeV}$ (the on-shell W/Z case), $300~\rm{GeV}$ for a positive $m_{\tilde{\chi}_1^0}$ and $ 35~{\rm GeV} \lesssim m_{\tilde{\chi}_2^0} - m_{\tilde{\chi}_1^0} \lesssim 90~{\rm GeV}$ (the off-shell W/Z case), and $220~\rm{GeV}$ when $ m_{\tilde{\chi}_2^0} - m_{\tilde{\chi}_1^0} = 15~{\rm GeV}$ (the extreme off-shell W/Z case). By contrast, $\tilde{\chi}_2^0$ is excluded only up to $210~\rm{ GeV}$ in mass for the off-shell W/Z case of the higgsino scenario, which occurs when
    $ m_{\tilde{\chi}_2^0} - m_{\tilde{\chi}_1^0} = 10~{\rm GeV}$ or  $ m_{\tilde{\chi}_2^0} - m_{\tilde{\chi}_1^0} \gtrsim 35~{\rm GeV}$.

\item \texttt{ATLAS-1911-12606~\cite{ATLAS:2019lng}}: Concentrate on compressed mass spectra case and search for electroweakino pair or slepton pair production, with two leptons and missing transverse momentum as the final state. The results are projected onto $\Delta m-\tilde{\chi}_2^0$ plane where $\Delta m \equiv m_{\tilde{\chi}_2^0} -  m_{\tilde{\chi}_1^0}$ for the electroweakino production. It is found that the tightest bound on higgsino-dominated $\tilde{\chi}_2^0$ is $193~{\rm GeV}$ in mass for $\Delta m \simeq 9.3~{\rm GeV}$, and the optimum bound on wino-dominated $\tilde{\chi}_2^0$ is $240~{\rm GeV}$ in mass when $\Delta m \simeq 7~{\rm GeV}$. Similarly, it is found that light-flavor sleptons should be heavier than about 250 GeV for $\Delta m_{\tilde{\ell}} = 10~{\rm GeV}$, where $m_{\tilde{\ell}} \equiv m_{\tilde{\ell}} - m_{\tilde{\chi}_1^0}$.

\item \texttt{CMS-SUS-20-001~\cite{CMS:2020bfa}}: Search for SUSY signal containing two oppositely charged same-flavor leptons and missing transverse momentum. This analysis studies not only strong sparticle productions, but also the electroweakino productions. The lepton originates from an on-shell or off-shell $Z$ boson in the decay chain, or from the decay of the produced sleptons. For the electroweakino pair production, the wino-dominated chargino and neutralino are explored up to $750~{\rm GeV}$ and $800~{\rm GeV}$, respectively, in mass. For the slepton pair production, the first two-generation sleptons are explored up to $700~{\rm GeV}$ in mass.

\end{itemize}

\begin{table}[tpb]
	\centering
	\caption{\label{Hadronic-R-values} Values of $R \equiv max\{S_i/S_{i,obs}^{95}\}$ for the analyses of the process $p p \to \tilde{\chi}_2^0 \tilde{\chi}_1^\pm \to 2 \tilde{\chi}_1^0 + WZ$ using the leptonic and hadronic signals of the vector bosons~\cite{ATLAS:2021moa,ATLAS:2021yqv}, respectively. SUSY benchmark points in wino simplified model are considered~\cite{ATLAS:2021yqv}. $R > 1 $ means that the point is experimentally excluded if the uncertainties involved in the calculation are neglected~\cite{Cao:2021tuh}, while $R < 1$ implies that the point is consistent with the experimental analysis.  Both analyses have exclusion capabilities on SUSY parameter space only when $m_{\tilde{\chi}_1^0} \lesssim 300~{\rm GeV}$, which is exhibited in Fig. 15(b) of Ref.~\cite{ATLAS:2021yqv}. }
	
\hspace{0.3cm}

	\begin{tabular}{l|ccc|ccc}
		\hline
		& \multicolumn{3}{c|}{Analysis in Ref.~\cite{ATLAS:2021moa}} & \multicolumn{3}{c}{Analysis in Ref.~\cite{ATLAS:2021yqv}} \\\hline
		\diagbox{$m_{\tilde{\chi}_1^0}~\rm{(GeV)}$}{R}{$m_{\tilde{\chi}_2^0/\tilde{\chi}_1^{\pm}}~\rm{(GeV)}$}&500&600&800&500&600&800\\\hline
		150 & 1.70& 1.07&0.39&0.19& 1.14& 1.45      \\
		200 & 1.54& 0.98&0.38& 0.11& 0.85& 1.52       \\
		250 & 1.28& 1.01&0.34& 0.10& 0.57& 1.19       \\
		300 & 0.96& 0.88&0.31& 0.11& 0.37& 1.04       \\\hline
	\end{tabular}
\end{table}

It should be noted that although the first reports are based on the analyses of $36~{\rm fb}^{-1}$ data, their exclusion capabilities are comparable with those of the second and fourth ones, which study the same signals but utilize $139~{\rm fb}^{-1}$ data. It should also be noted that the search for charginos and neutralinos using fully hadronic final states of W/Z and Higgs bosons with $139~{\rm fb}^{-1}$ data~\cite{ATLAS:2021yqv} is not considered.  As shown in Fig. 12(c) of Ref.~\cite{ATLAS:2021yqv}, this analysis excludes a wino mass up to 1060 GeV when $m_{\tilde{\chi}_1^0}$ is below 400 GeV, and the mass splitting between the winos and $\tilde{\chi}_1^0$ is larger than about 400 GeV. It seems much stronger than the second report in limiting SUSY parameter space.  This conclusion, however, can only be applied indirectly to this study due to the following facts. First, the results in the Fig. 12(c) are based on  the assumption $Br(\tilde{\chi}_2^0 \to \tilde{\chi}_1^0 Z) = Br(\tilde{\chi}_2^0 \to \tilde{\chi}_1^0 h) = 50\%$ and the signals from the processes $p p \to \tilde{\chi}_2^0 \tilde{\chi}_1^\pm, \tilde{\chi}_1^\pm \tilde{\chi}_1^\mp \to 2 \tilde{\chi}_1^0 + WW/WZ/Wh$ in the wino scenario. The most significant contribution to $R \equiv max\{S_i/S_{i,obs}^{95}\}$, where $S_i$ denotes the simulated event number of the $i$-th signal region (SR) in the analysis, and $S_{i,obs}^{95}$ represents
its corresponding $95\%$ confidence level upper limit,  comes from the process $p p \to \tilde{\chi}_2^0 \tilde{\chi}_1^\pm \to 2 \tilde{\chi}_1^0 + Wh$. It can exclude $m_{\tilde{\chi}_2^0/\tilde{\chi}_1^\pm}$ up to $1060~{\rm GeV}$ for massless $\tilde{\chi}_1^0$ and up to $950~{\rm GeV}$ for $m_{\tilde{\chi}_1^0} = 400~{\rm GeV}$ if $Br(\tilde{\chi}_2^0 \to \tilde{\chi}_1^0 h) = 100 \%$ is assumed. In contrast, the minimum contribution arises from the signal $p p \to \tilde{\chi}_1^\pm \tilde{\chi}_1^\mp \to 2 \tilde{\chi}_1^0 + W^\pm W^\mp$, which can only exclude a small region on $m_{\tilde{\chi}_1^\pm}-m_{\tilde{\chi}_1^0}$ plane, characterized by $m_{\tilde{\chi}_1^0} \lesssim 90~{\rm GeV}$ and $640~{\rm GeV} \lesssim m_{\tilde{\chi}_1^\pm} \lesssim 750~{\rm GeV}$, for $Br(\tilde{\chi}_1^\pm \to \tilde{\chi}_1^0 W^\pm) = 100\%$. These conclusions can be inferred from Fig. 15 of the report. Evidently, smaller  $Br(\tilde{\chi}_2^0 \to \tilde{\chi}_1^0 h)$ and $Br(\tilde{\chi}_2^0 \to \tilde{\chi}_1^0 Z)$ will weaken significantly the exclusion capability in the Fig. 12(c). Second, the analyses of the process $p p \to \tilde{\chi}_2^0 \tilde{\chi}_1^\pm \to 2 \tilde{\chi}_1^0 + WZ$ using the leptonic signal~\cite{ATLAS:2021moa} and the hadronic signal~\cite{ATLAS:2021yqv}, respectively, are compared in Fig.15(b) of Ref.~\cite{ATLAS:2021yqv}. We extract $R$ values from this figure for some parameter points in Table~\ref{Hadronic-R-values}.
This table indicates that the hadronic analysis is more restrictive than the leptonic analysis only when $m_{\tilde{\chi}_2^0/\tilde{\chi}_1^\pm} \gtrsim 600~{\rm GeV}$ for the same $m_{\tilde{\chi}_1^0}$ with $m_{\tilde{\chi}_1^0} \lesssim 300~{\rm GeV}$ in the wino simplified model. Similar conclusion can be acquired for the $Wh$ signal by analyzing the process $p p \to \tilde{\chi}_2^0 \tilde{\chi}_1^\pm \to 2 \tilde{\chi}_1^0 + Wh$ with the leptonic and hadronic signals~\cite{ATLAS:2021yqv,ATLAS:2020pgy}, respectively, which is shown in Fig. 15(c) of Ref.~\cite{ATLAS:2021yqv}. This remarkable characteristic comes from the fact that the hadronic analysis rejects large backgrounds by identifying high-$p_T$ bosons using large-radius jets and jet substructure information, and this strategy is inefficient for low-$p_T$ bosons. Finally, we emphasize that one of the main aims of this study is to explain the muon g-2 anomaly. Owing to the sparticle mass spectra preferred by the anomaly, the branching ratios of the wino-like particles decaying into W, Z and Higgs bosons can be suppressed significantly, and more importantly, the $p_T$s of the bosons are generally softened when compared with the simplified model predictions in Ref.~\cite{ATLAS:2021yqv}. As a result, the hadronic analysis is scarcely crucial in this study. We will return to this issue in Section 3.5.

In summary, the GNMSSM has the following characteristics:
\begin{itemize}
\item $\tan \beta$, $M_1$, $M_2$, $\mu_{tot}$, $\tilde{m}_{\tilde{\mu}_L}$, and $\tilde{m}_{\tilde{\mu}_R}$: They determine the magnitude of $a_\mu^{\rm SUSY}$. In particular, $\tan \beta$, $M_2$, $\mu_{tot}$, and $\tilde{m}_{\tilde{\mu}_L}$ play a crucial role in explaining the muon g-2 anomaly, and the dimensional parameters have been restricted significantly by the LHC search for SUSY.
\item $A_\lambda$ and $m_3$: They determine the mass of the heavy doublet-dominated scalars,
but hardly affect the other Higgs bosons' property if $\lambda$ is small.

\item $\mu_{eff}$, $A_\kappa$, $\mu^\prime$, and $m_S^\prime$: They are parameters describing the properties of the singlet field $S$ and thus are scarcely restricted by the LHC experiments. Since  $m_{\tilde{\chi}_1^0}$, $m_{h_s}^2$, and $m_{A_s}^2$ depend on parameters $\mu^\prime$,  $A_\kappa$, and $m_S^{\prime 2}$ linearly in the small $\lambda$ limit, these physical masses are not correlated and may vary freely within certain ranges.

\item $\lambda$ and $\kappa$: They are Yukawa couplings in the Higgs sector. Once the singlino-dominated $\tilde{\chi}_1^0$, $h_s$, and $A_s$ form a secluded DM sector, they determine the magnitudes of the DM-nucleon scattering and the relic density, respectively.
\item $m_{\tilde{t}_L}$, $m_{\tilde{t}_R}$, and $A_t$: They affect the mass of the SM-like Higgs boson significantly  by radiative corrections.
\end{itemize}

\begin{table}[tbp]
\caption{Parameter space explored in this study. Other dimensional parameters not crucial to this study are fixed at 3 TeV, including the SUSY parameters for
the first and third generation sleptons, three generation squarks (except for $A_t$ and $A_b$ with the assumption $A_b = A_t$), and gluinos. All the parameters are defined at the renormalization scale $Q=1~{\rm TeV}$. \label{Table3}}
\centering

\vspace{0.3cm}

\resizebox{0.7\textwidth}{!}{
\begin{tabular}{c|c|c|c|c|c}
\hline
Parameter & Prior & Range & Parameter & Prior & Range   \\
\hline
$\lambda$ & Flat & $0 \sim0.7$ & $\kappa$ & Flat & $-0.7 \sim 0.7$  \\
$\tan \beta$ & Flat & $1 \sim60$ &$A_{\kappa}/{\rm TeV}$ & Flat & $-1.0 \sim 1.0 $ \\
$\mu_{\rm tot}/{\rm TeV}$ & Log & $0.1\sim 1.0$ &$\mu_{\rm eff}/{\rm TeV}$ & Log & $10^{-3}\sim 1.0$ \\
$\mu^\prime/{\rm TeV}$ & Flat & $-1.0\sim 1.0$ & $A_t/{\rm TeV}$ & Flat & $-5.0\sim 5.0$ \\
$M_1/{\rm TeV}$ & Flat & $-1.5\sim1.5$ & $M_2/{\rm TeV}$ & Log & $0.1\sim 1.5$ \\
$\tilde{m}_{\tilde{\mu}_L}/{\rm TeV}$ & Log & $0.1\sim 1.0$ &$\tilde{m}_{\tilde{\mu}_R}/{\rm TeV}$ & Log & $0.1\sim 1.0$ \\
$m_S^{\prime 2}/{\rm TeV^2}$ & Flat & $- 1.0\sim 1.0 $& & &  \\
\hline
\end{tabular}}
\end{table}

%\vspace{-0.7cm}

\section{\label{numerical study}Status of singlino-dominated DM}

This study utilizes the package \texttt{SARAH-4.14.3}~\cite{Staub:2008uz, Staub:2012pb, Staub:2013tta, Staub:2015kfa} to build the model file of the GNMSSM,
the codes \texttt{SPheno-4.0.4}~\cite{Porod:2003um, Porod:2011nf} and \texttt{FlavorKit}~\cite{Porod:2014xia} to generate particle mass spectra and compute low energy
observables, such as $a_\mu^{\rm SUSY}$ and B-physics observables,  and the package \texttt{MicrOMEGAs-5.0.4}~\cite{Belanger:2001fz, Belanger:2005kh, Belanger:2006is, Belanger:2010pz, Belanger:2013oya, Barducci:2016pcb} to calculate DM observables, assuming that the lightest neutralino is the sole DM candidate in the universe. Bounds from the direct search for extra Higgs bosons at LEP, Tevatron, and LHC and the fit of $h$'s property to LHC Higgs data are implemented by the programmes~\texttt{HiggsBounds-5.3.2}~\cite{HB2008jh,HB2011sb,HB2013wla,HB2020pkv} and \texttt{HiggsSignal-2.2.3}~\cite{HS2013xfa,HSConstraining2013hwa,HS2014ewa,HS2020uwn}, respectively.

%\vspace{-0.2cm}

\subsection{\label{scan}Research strategy}

\begin{figure}[t]
	\centering
	\includegraphics[width=0.45\textwidth]{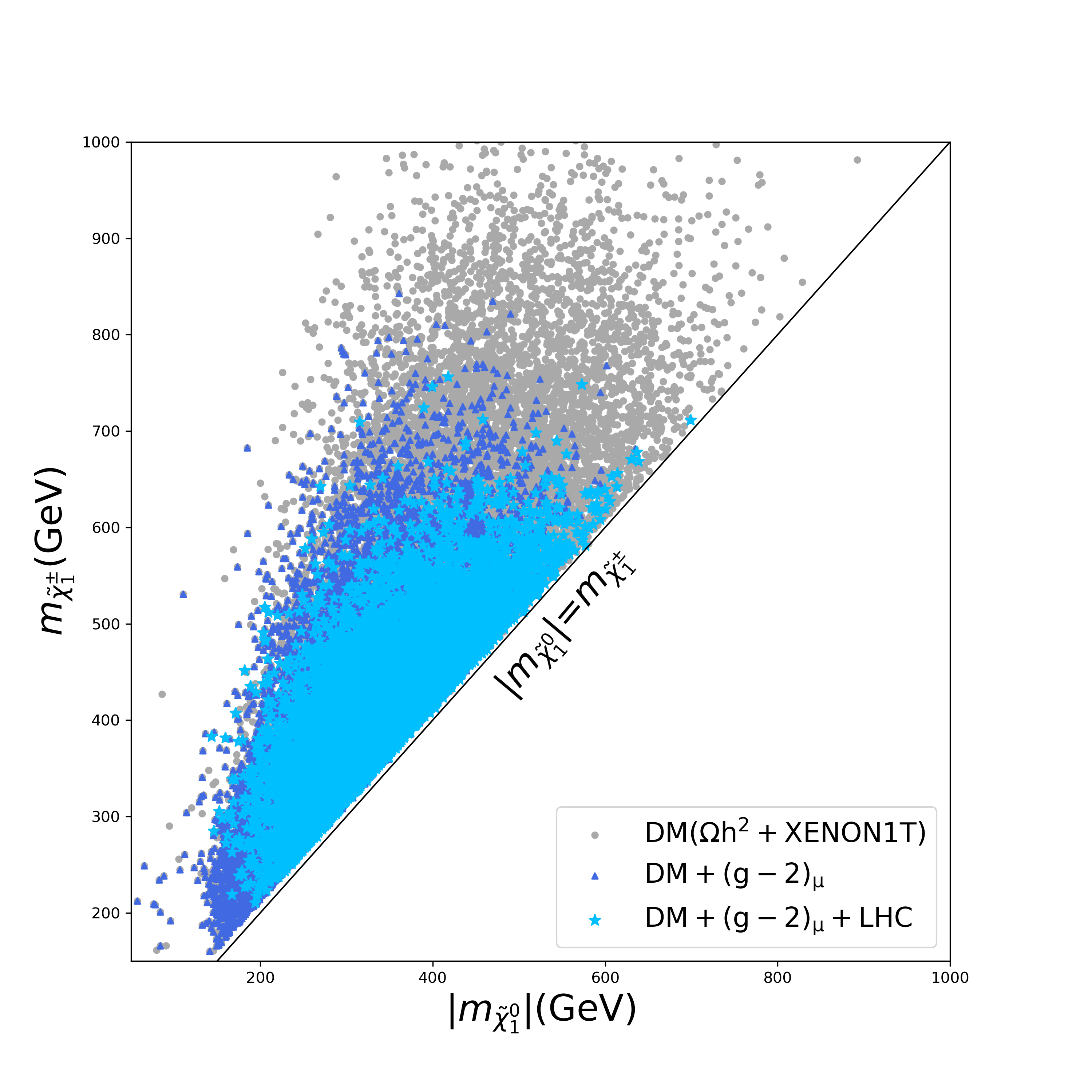}\hspace{-0.3cm}
	\includegraphics[width=0.45\textwidth]{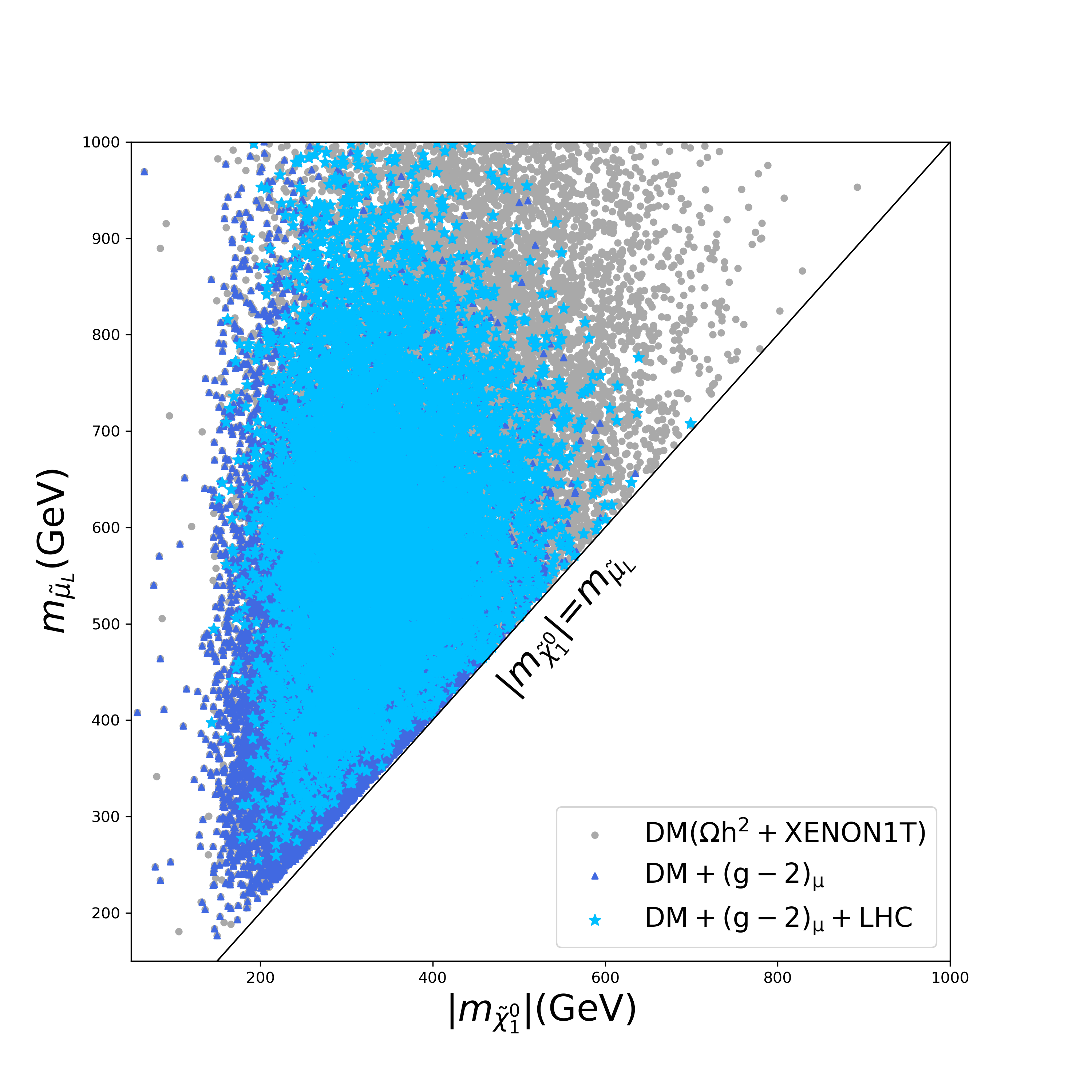}%\hspace{-0.3cm}

\vspace{-0.5cm}

	\caption{\label{fig1} Projection of the obtained samples onto the $|m_{\tilde{\chi}_1^0}|-m_{\tilde{\chi}_1^\pm}$ plane~(left panel) and the $|m_{\tilde{\chi}_1^0}|-m_{\tilde{\mu}_L}$ plane~(right panel). The dark grey points denote the samples that satisfy all constraints listed in the text, in particular those from the DM experiments. The royal blue triangles represent those that can further explain the muon g-2 anomaly at $2\sigma$ level, and the sky blue stars are part of the royal blue triangles which agree with the results from the LHC search for SUSY. }
\end{figure}

The \texttt{MultiNest} algorithm~\cite{Feroz:2008xx} is employed to scan comprehensively the parameter space in Table~\ref{Table3}. The $n_{\rm live}$ parameter in the algorithm controls the number of active points sampled in each iteration of the scan, and $n_{\rm live} = 8000$ is set. The following likelihood function is constructed to guide the scan:
\begin{eqnarray}
\mathcal{L} = \mathcal{L}_{a_\mu} \times \mathcal{L}_{const}, \label{likelihood}
\end{eqnarray}
where $\mathcal{L}_{a_\mu}$ is the likelihood function of the muon $g-2$ anomaly given by
\begin{eqnarray}
\mathcal{L}_{a_\mu} \equiv Exp\left[-\frac{1}{2} \left( \frac{a_{\mu}^{\rm SUSY}- \Delta a_\mu}{\delta a_\mu }\right)^2\right] = Exp\left[-\frac{1}{2} \left( \frac{a_{\mu}^{\rm SUSY}- 2.51\times 10^{-9}}{5.9\times 10^{-10} }\right)^2\right], \nonumber
\end{eqnarray}
with $\Delta a_\mu \equiv a_\mu^{\rm Exp} - a_\mu^{\rm SM}$ and $\delta a_\mu$ denoting the difference between the experimental central value of $a_\mu$ and its SM prediction, and  the total uncertainties in determining $\Delta a_\mu$, respectively~\cite{Abi:2021gix,Bennett:2006fi,Aoyama:2020ynm,Aoyama:2012wk,Aoyama:2019ryr,Czarnecki:2002nt,Gnendiger:2013pva,Davier:2017zfy,
Keshavarzi:2018mgv,Colangelo:2018mtw,Hoferichter:2019gzf,Davier:2019can,Keshavarzi:2019abf,Kurz:2014wya,Melnikov:2003xd,Masjuan:2017tvw,
Colangelo:2017fiz,Hoferichter:2018kwz,Gerardin:2019vio,Bijnens:2019ghy,Colangelo:2019uex,Blum:2019ugy,Colangelo:2014qya}. $\mathcal{L}_{const}$ represents the impacts of experimental constraints on the theory:  $\mathcal{L}_{const} = 1$ by our definition if the constraints are satisfied and $\mathcal{L}_{const} = Exp[-100]$ if they are not. These constraints include $2\sigma$ bounds of the DM relic density~\cite{Planck:2018vyg}, upper bounds of the XENON-1T experiment on the DM-nucleon scattering~\cite{Aprile:2018dbl,Aprile:2019dbj}, consistence of $h$'s property with the LHC Higgs data at $95\%$ confidence level~\cite{HS2020uwn}, collider searches for extra Higgs bosons~\cite{HB2020pkv}, $2\sigma$ bounds on the branching ratios of $B \to X_s \gamma$ and $B_s \to \mu^+ \mu^-$~\cite{PhysRevD.98.030001}, and the vacuum stability of the scalar potential consisting of the Higgs fields and the last two generation slepton fields~\cite{Camargo-Molina:2013qva,Camargo-Molina:2014pwa}. In Refs.~\cite{Cao:2021tuh,Cao:2022chy}, we presented the details of these constraints. We emphasize one subtle thing about the prior distributions: if a log distribution is adopted for $\mu_{eff}$, more than $80\%$ of the samples obtained in the scan predict a singlino-dominated $\tilde{\chi}_1^0$, while if a flat distribution is adopted, nearly all the samples correspond to a bino-dominated $\tilde{\chi}_1^0$.
The reason is as follows: a small $\lambda$ of ${\cal{O}}(10^{-2})$ is favorite by the singlino-dominated DM to suppress the DM-nucleon scattering, and given $\mu_{eff} \equiv \lambda v_s/\sqrt{2}$ with $v_s \sim {\cal{O}}(1~\rm TeV)$, the typical size of $\mu_{eff}$ is around several tens of GeV.
We verified that there is no such a phenomenon for the other parameters. We also verified that if one replaces the log distributions in Table~\ref{Table3} by flat distributions and the region of $\mu_{eff} \leq 1~{\rm TeV}$ by $\mu_{eff} \leq 30~{\rm GeV}$, and performs another scan of the parameter space, he can reproduce the crucial characteristics of the dark gray and royal blue points in Figs.~\ref{fig1} and \ref{fig2}.

Next, only the samples predicting a singlino-dominated DM and satisfying all the constraints are investigated. The following processes are studied to decide whether they pass the constraints from the LHC search for SUSY or not:
\begin{eqnarray}
pp &\to& \tilde{\chi}_i^0\tilde{\chi}_j^{\pm}, \quad i = 2, 3, 4, 5, \quad j = 1, 2; \\
pp &\to& \tilde{\chi}_i^{\pm}\tilde{\chi}_j^{\mp}, \quad i,j = 1, 2; \\
pp &\to& \tilde{\chi}_i^{0}\tilde{\chi}_j^{0}, \quad i,j = 2, 3, 4, 5; \\
pp &\to& \tilde{\mu}_i^\ast \tilde{\mu}_j,\quad i,j = 1, 2; \\
pp &\to& \tilde{\nu}_{\mu}^\ast \tilde{\nu}_\mu.
\end{eqnarray}
Specifically, the cross-sections of these processes at $\sqrt{s}$ = 13 TeV are calculated at the next-to-leading order (NLO) by the package \texttt{Prospino2}~\cite{Beenakker:1996ed}. In order to save computing time, the program \texttt{SModelS-2.1.1}~\cite{Khosa:2020zar} is initially used to exclude the obtained samples. Given that this program's capability to implement the LHC constraints is limited by its database and strict working prerequisites, the surviving samples are further surveyed by simulating the analyses in Tables~\ref{Table1} and~\ref{Table2}. This study is accomplished by the following steps: 60000 and 40000 events are generated for the electroweakino and slepton production processes, respectively, by package \texttt{MadGraph\_aMC@NLO}~\cite{Alwall:2011uj, Conte:2012fm}; relevant parton shower and hadronization are finished by program \texttt{PYTHIA8}~\cite{Sjostrand:2014zea}; the resulting event files are then fed into package \texttt{CheckMATE-2.0.29}~\cite{Drees:2013wra,Dercks:2016npn, Kim:2015wza} to calculate the $R$-value defined by $R \equiv max\{S_i/S_{i,obs}^{95}\}$, where $S_i$ denotes the simulated event number of the $i$-th SR in the analyses of Tables~\ref{Table1} and~\ref{Table2}, and $S_{i,obs}^{95}$ represents its corresponding $95\%$ confidence level upper limit. Note that program \texttt{Delphes} is encoded in the CheckMATE for detector simulation~\cite{deFavereau:2013fsa}.

\subsection{\label{region}Key features of the results}

\begin{table}[tpb]
	\centering
	\caption{\label{Table4} Number of the samples obtained in the scan, which are categorized by the DM dominant annihilation mechanisms and whether the LHC restrictions are considered or not. The total numbers of the royal blue triangles and the sky blue stars in Fig.~\ref{fig1} are 20889 and 8517, respectively, which are listed on the first row. }

\hspace{0.1cm}

	\begin{tabular}{l|c|c}
		\hline
		Annihilation Mechanisms& Without LHC Constraints &  With LHC Constraints       \\ \hline
		\multicolumn{1}{l|}{Total Samples}                                                                        &20889             &8517\\
		 {$\tilde{\chi}_1^0\tilde{\chi}_1^0 \rightarrow h_{s}h_{s}$}                                                                 &10925              &4028\\
		\multicolumn{1}{l|}{$\tilde{\chi}_1^0\tilde{\chi}_1^0 \rightarrow h_{s}A_{s}$}                                                                 &8792                 &4115\\
		\multicolumn{1}{l|}{$\tilde{\chi}_1^0\tilde{\chi}_1^0 \rightarrow hA_{s}$}
		&66              &18\\
		%\multicolumn{1}{l|}{$A_{s}$-funnel}                                          &47              &9\\
		\multicolumn{1}{l|}{$A_{s}$-funnel}                                          &310              &62\\
		\multicolumn{1}{l|}{$\tilde{\chi}_1^0\tilde{\chi}_1^0 \rightarrow A_{s}A_{s}$}                                                                 &6              &0\\	
		\multicolumn{1}{l|}{$\tilde{\chi}_1^0\tilde{\chi}_1^0 \rightarrow hh_{s}$}
		&2              &0\\
		\multicolumn{1}{l|}{Higgsino co-annihilation}                                                                    &385              &227 \\
		\multicolumn{1}{l|}{Wino co-annihilation}                                                                    &85              &61 \\
		\multicolumn{1}{l|}{$\tilde{\nu}_\mu/\tilde{\mu}_L$ co-annihilation}                                                                    &286              &5 \\
		\multicolumn{1}{l|}{$\tilde{\mu}_R$ co-annihilation}                                                                    &32              &1 \\
		\multicolumn{1}{l|}{$h_s-$ or $A_s-$funnel}                                                                    &0              &0 \\
		\multicolumn{1}{l|}{$Z-$ or $h-$funnel}                                                                    &0              &0 \\
\hline
		\end{tabular}
\end{table}

\begin{figure}[t]
	\centering
	\includegraphics[width=0.45\textwidth]{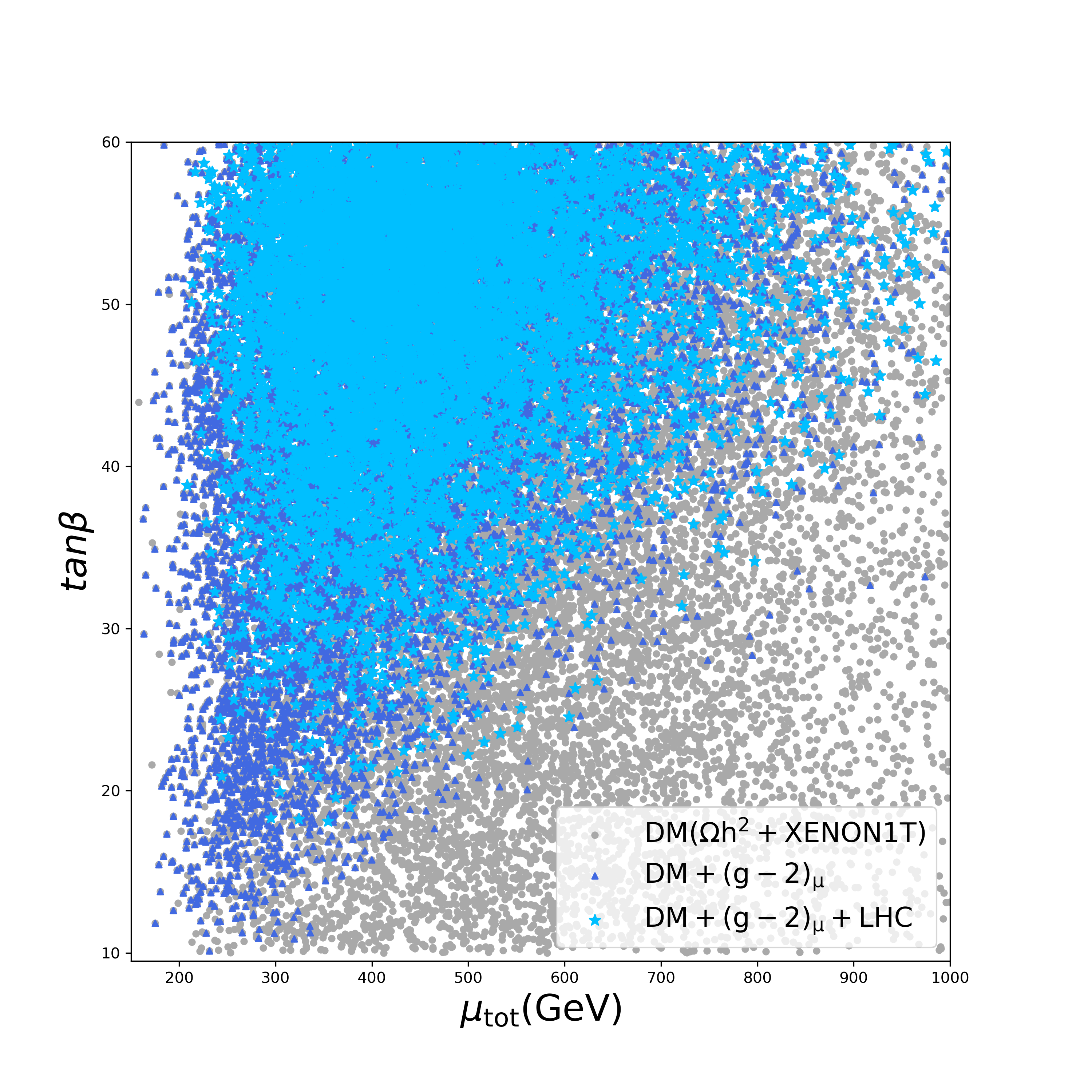}\hspace{-0.3cm}
	\includegraphics[width=0.45\textwidth]{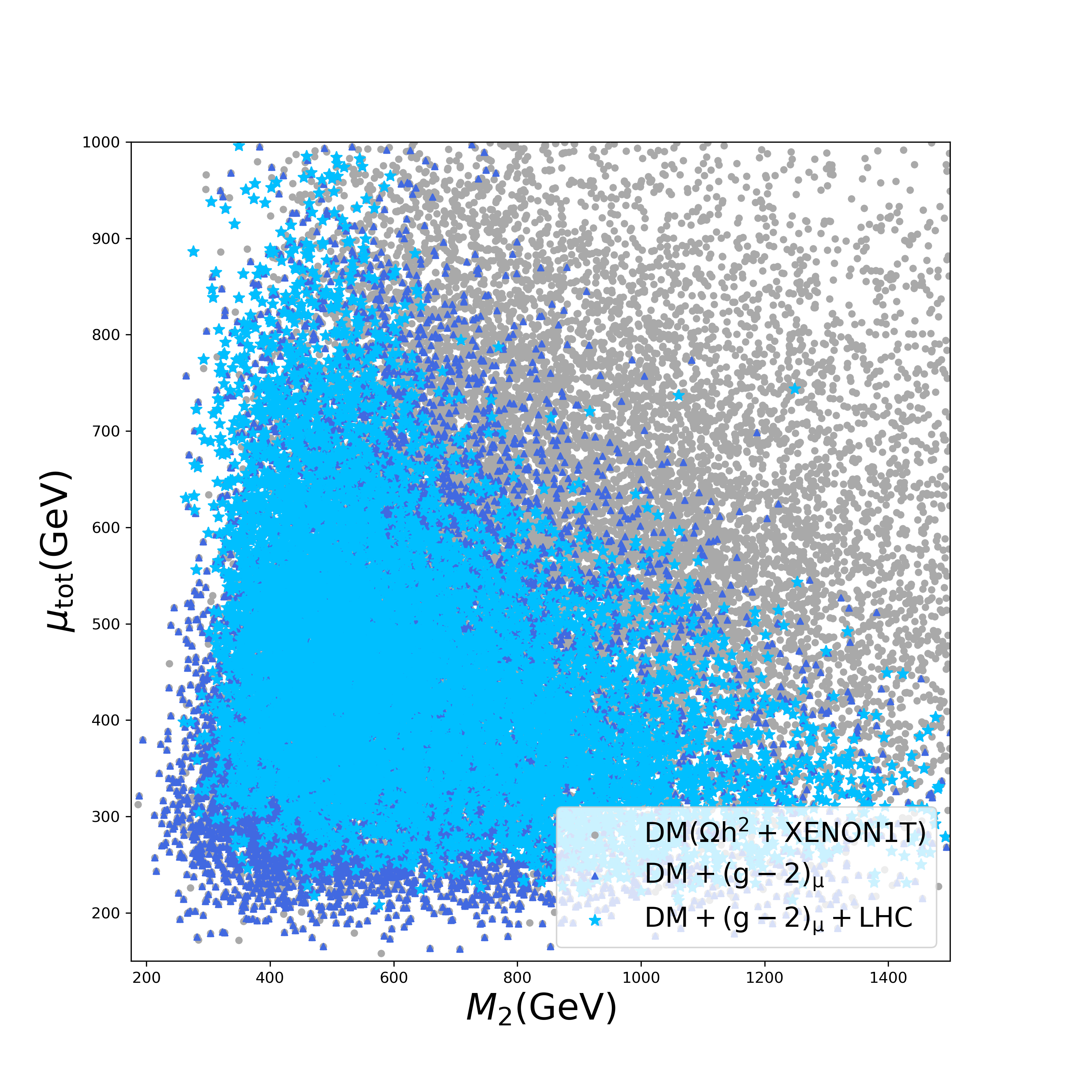} %\hspace{-0.3cm}
	\includegraphics[width=0.45\textwidth]{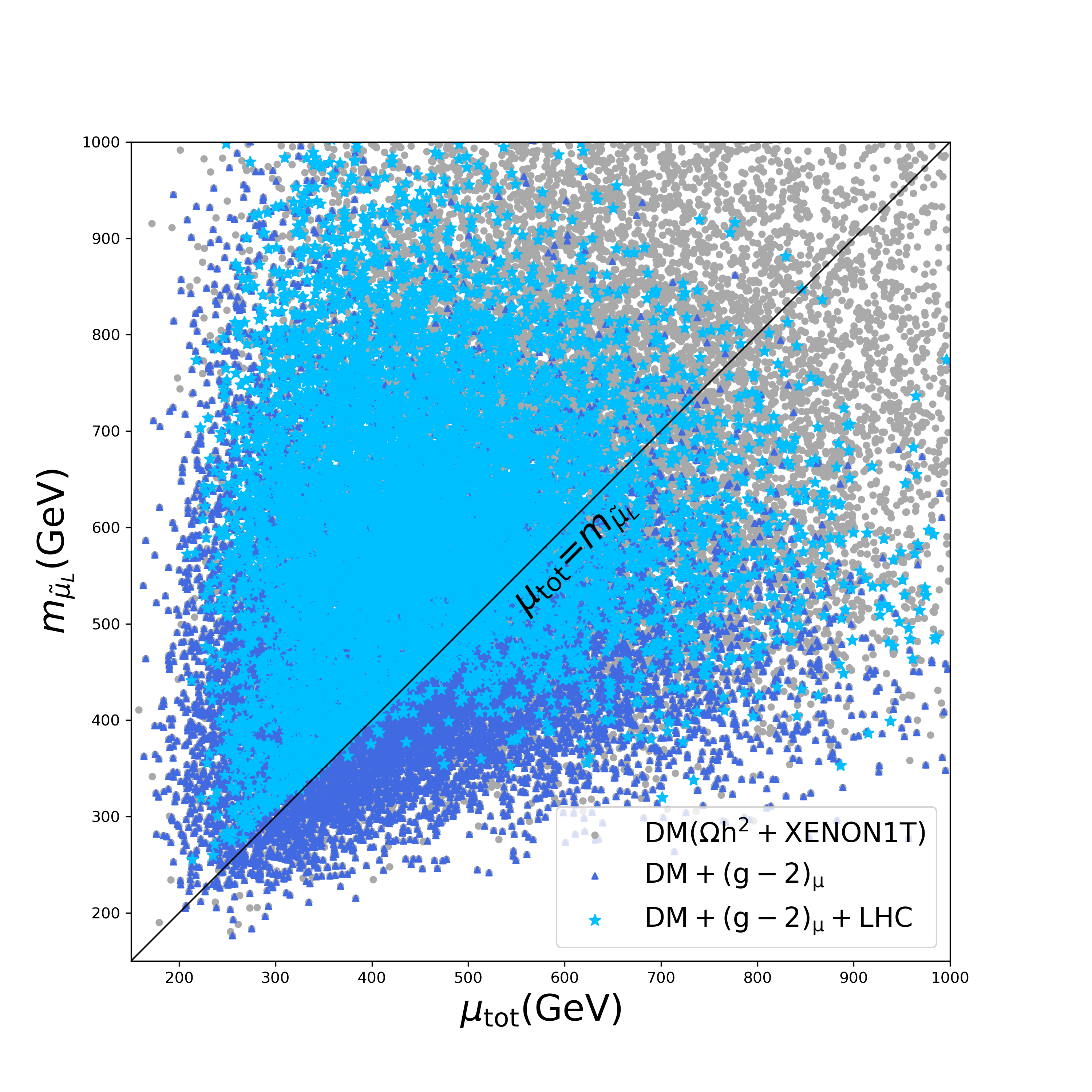}\hspace{-0.3cm}
	\includegraphics[width=0.45\textwidth]{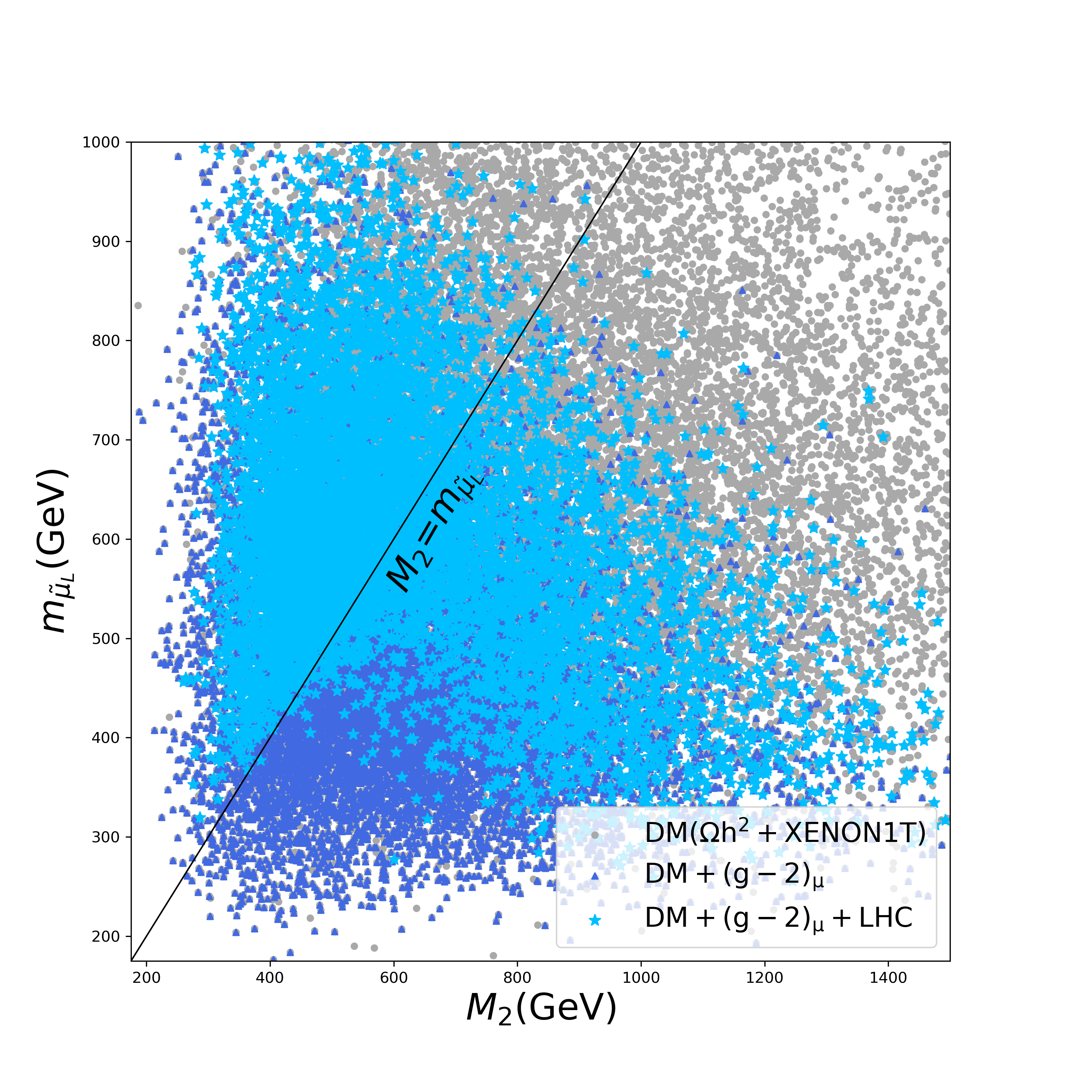}%\hspace{-0.3cm}\\

\vspace{-0.5cm}

	\caption{\label{fig2}
		Similar to Fig.~\ref{fig1}, but showing the correlations of the parameters that the muon g-2 is sensitive to. }
\end{figure}

We first study the DM physics of the GNMSSM comparatively. In the $Z_3$-NMSSM, the DM candidate is bino-dominated in most cases, and it obtains the measured relic density mainly by co-annihilating with wino-dominated electroweakinos or the sleptons with the left-handed chirality as their dominant components. This conclusion is reflected in Figs.~2 and 3 of Ref.~\cite{Cao:2022htd}, where we studied the $Z_3$-NMSSM similarly to this work, and projected the obtained samples onto $|m_{\tilde{\chi}_1^0}|-m_{\tilde{\chi}_1^\pm}$ and $|m_{\tilde{\chi}_1^0}|-m_{\tilde{\mu}_L}$ planes ($m_{\tilde{\chi}_1^\pm}$ and $m_{\tilde{\mu}_L}$ denote the mass of the lightest chargino and the left-handed dominant smuon, respectively)\footnote{In Bayesian statistics, the parameter space of new physics theory is described by the posterior probability density function (PDF), which is determined by the likelihood function of the research object, the prior distributions of input parameters, and the considered parameter space~\cite{Jeffreys:1939xee}. In the simplest case, where all the inputs are initially assumed to be flatly distributed, the posterior PDF is proportional to the likelihood function and achieves its maximum when the theory predicts the central values of experimental measurements. For the scan done in the last subsection, its sample number within a hypercube of the parameter space is proportional to the hypercube's Bayesian evidence, acquired by integrating the PDF over the hypercube~\cite{Jeffreys:1939xee}. So, if the PDF of a scenario is sizable only within very narrow corners of the parameter space, it is hardly encountered in the scan by the MultiNest algorithm, and its corresponding samples are rare~\cite{Feroz:2008xx}. The fundamental reason is that the parameters of this scenario need to be finely tuned to meet relevant experimental results. In the $Z_3$-NMSSM, DM may be singlino-dominated, and it achieves the measured density either through co-annihilating with the higgsino-dominated electroweakinos or through the $h_s$- or $A_s$-funnel, as we recapitulated in Sec.~\ref{Introduction}. In addition, the bino-dominated DM in both MSSM and $Z_3$-NMSSM may achieve the measured density by $Z$- or $h$-funnel~\cite{Calibbi:2013poa,Belanger:2013pna,Hamaguchi:2015rxa,
Cao:2015efs,Barman:2017swy,Wang:2020dtb,KumarBarman:2020ylm,Abdallah:2020yag,
VanBeekveld:2021tgn,Barman:2022jdg}. These scenarios, however, become more and more finely tuned after considering the improved DM direct detection experiments, which we explained in Sec.~\ref{Introduction}. They are usually missed in the scan and, thus, of less interest when one aims to capture the global characteristics of the $Z_3$-NMSSM. }.
Here we project the samples of this study onto the same planes to obtain Fig.~\ref{fig1}. Evidently, this figure differs significantly from the corresponding ones in Ref.~\cite{Cao:2022htd} by that only a small portion of the samples predict $m_{\tilde{\chi}_1^\pm} \simeq |m_{\tilde{\chi}_1^0}|$ or $m_{\tilde{\mu}_L} \simeq |m_{\tilde{\chi}_1^0}|$, which is the necessary condition for the co-annihilations. To further clarify this point, we categorize the obtained samples by their dominant annihilation mechanisms and count the number of samples in each category. We present the results in Table~\ref{Table4}. This table shows that for most of the samples, the DM annihilates mainly by $\tilde{\chi}_1^0 \tilde{\chi}_1^0 \to h_s h_s, h_s A_s$ to obtain the measured relic density, which coincides with the analyses in Sec.~\ref{theory-section}.  Fig.~\ref{fig1} and Table~\ref{Table4} also indicate that the LHC constraints are rather strong and they exclude about $59\%$ of the obtained samples. As a result, the lower bound of $|m_{\tilde{\chi}_1^0}|$ for the samples is lifted from about $50~{\rm GeV}$ to about $140~{\rm GeV}$ by the constraints\footnote{As explained in footnote 6, the samples obtained in this study depend strongly on the likelihood function in Eq.(\ref{likelihood}). The statement in the text is based on the assumption that the DM is fully responsible for the measured density. If it is relaxed by setting an upper bound on the density, more parameter spaces will open, and the GNMSSM's phenomenology becomes more complicated. In this case, the lower bound of $|m_{\tilde{\chi}_1^0}|$ may be significantly reduced, and its exact value is determined only through a renewed study. This issue, however, is beyond the scope of this study. In addition, we remind that the same arguments can be applied to the conclusions in Ref.~\cite{Cao:2022htd}. }. By contrast, we concluded in~\cite{Cao:2022htd} that about $66\%$ of the obtained samples are excluded if the bino-dominated DM in the $Z_3$-NMSSM is concerned, and $|m_{\tilde{\chi}_1^0}|$ must be larger than about $260~{\rm GeV}$ after considering the constraints.
The fundamental reason for these phenomena is at least two-fold. One is that explaining the muon g-2 anomaly requires more than one sparticle to be moderately light~\cite{Chakraborti:2020vjp}, which can strengthen the SUSY signals at the LHC. The other is as the DM becomes lighter, more missing transverse energy is to be emitted from the sparticle productions at the LHC, which can improve the exclusion capability of the experimental analyses.

Next, we focus on the explanation of the muon g-2 anomaly. Since the 'WHL' contribution dominates over the other ones, we show the correlations of any two of the three parameters, $\mu_{tot}$, $M_2$, and $m_{\tilde{\mu}_L}$ in Fig.~\ref{fig2}. We also plot the correlation between $\mu_{tot}$ and $\tan \beta$. This figure reveals the following facts:
\begin{itemize}
\item The muon g-2 anomaly and the observables in DM physics may prefer different parameter spaces of the GNMSSM, but moderately broad parameter regions can still accommodate both. For example, if one requires the theory to explain the anomaly at $2\sigma$ confidence level, $\mu_{\rm tot}$, $M_2$, and $m_{\tilde{\mu}_L}$ must be less than about $1000~{\rm GeV}$, $1500~{\rm GeV}$, and $1000~{\rm GeV}$, respectively. These upper bounds depend strongly on the value of $\tan \beta$, e.g., they are about $350~{\rm GeV}$, $550~{\rm GeV}$, and $400~{\rm GeV}$, respectively, for $\tan \beta =10$, and become about $800~{\rm GeV}$, $1100~{\rm GeV}$, and $800~{\rm GeV}$ for $\tan \beta = 30$\footnote{Note that the bounds will shrink significantly if one requires the theory to explain the anomaly at $1\sigma$ level. In this case, we find $\mu_{tot} \lesssim 650~{\rm GeV}$, $M_2 \lesssim 900~{\rm GeV}$, and $m_{\tilde{\mu}_L} \lesssim 650~{\rm GeV}$ for $\tan \beta=30$.}. By contrast, the DM physics is not sensitive to the four parameters.

\item As any one of $\mu_{tot}$, $M_2$, and $m_{\tilde{\mu}_L}$ increases, the other two parameters tend to decrease. In other words, these three parameters can not be very large simultaneously in explaining the muon g-2 anomaly. This characteristic makes the LHC restrictions on the theory rather strong.

\item There are two cases that the LHC restrictions are particularly strong. One is characterized by $\tan \beta \lesssim 20$, where the winos, the higgsinos, and the left-handed dominant smuon are all lighter than about 500 GeV. This feature is shown in the top left panel for parameter $\mu_{tot}$. The other case is characterized by predicting a $\tilde{\mu}_L$  lighter than winos and/or higgsinos, where the heavy electroweakinos may decay into the slepton first and thus enhance the leptonic signal of the electroweakino pair production processes (compared with the case that $\tilde{\mu}_L$ is heavier than the electroweakinos). This feature accounts for the wedge-shaped region on $M_2-m_{\tilde{\mu}_L}$ plane, which is displayed by the LHC limits in the bottom right panel. We elaborate on this point by fixing $m_{\tilde{\mu}_L} = 350~{\rm GeV}$ in the following. For $M_2 \lesssim 350~{\rm GeV}$, although the wino pair production cross-sections exceed $300~{\rm fb}$~\cite{Fuks:2012qx,Fuks:2013vua}, there are still few samples surviving the LHC constraints. We verified that it is due to small mass splittings of the wino-like particles with $\tilde{\chi}_1^0$, where the wino-like particles decay into $\tilde{\chi}_1^0$ and a soft virtual Z or W.  In the case of $350~{\rm GeV} \lesssim M_2 \lesssim 600~{\rm GeV}$, all the samples are excluded by the SUSY searches. This is because the wino-like particles are copiously produced at the LHC since they are moderately light, and simultaneously the branching ratios of their decays into the slepton are sizable. Specifically, we find that the ratios are always larger than $10\%$ and may reach $60\%$ in the optimum cases. With the further increase of $M_2$, the wino pair production rates rapidly decrease. Consequently, the LHC constraints are weakened, which is reflected by the frequent appearance of sky blue points in the bottom right panel. We add that this discussion can be applied to the $\mu_{tot}-m_{\tilde{\mu}_L}$ plane in the bottom left panel.

\item The LHC restrictions have required $\tan \beta \gtrsim 18$, $\mu_{tot} \gtrsim 210~{\rm GeV}$, $M_2 \gtrsim 260~{\rm GeV}$, $m_{\tilde{\mu}_L} \gtrsim 255~{\rm GeV}$, and $m_{\tilde{\mu}_R} \gtrsim 240~{\rm GeV}$.  We emphasize that moderately light higgsinos are experimentally allowed in explaining the muon g-2 anomaly. This is one of the advantages of the GNMSSM over the MSSM and the $Z_3$-NMSSM.
\end{itemize}

\begin{figure}[t]
	\centering
	\includegraphics[height=7cm,width=12cm]{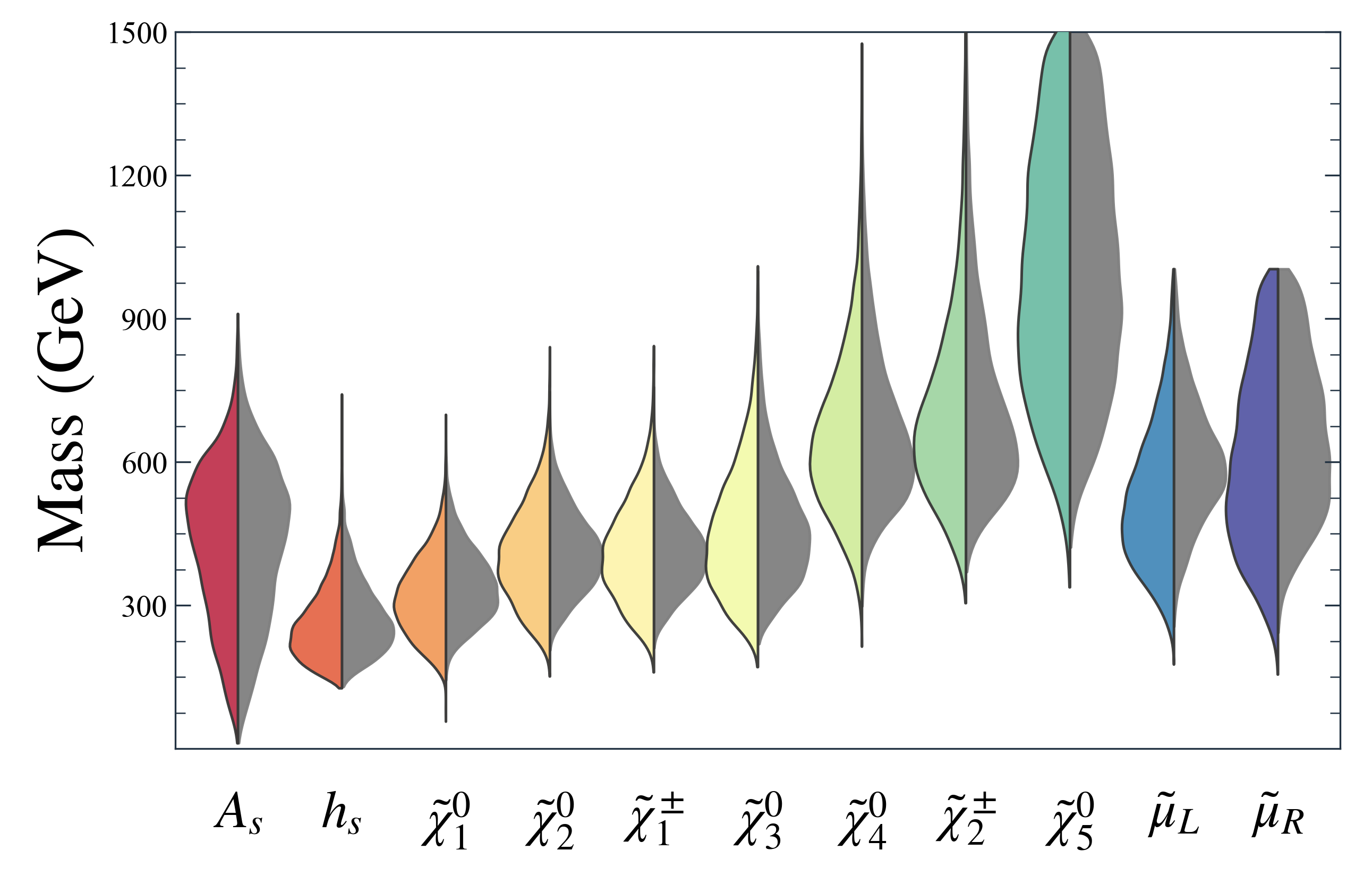} %\hspace{-0.3cm}

\vspace{-0.5cm}

	\caption{\label{fig3} Split violin plot showing the mass distributions of the particles beyond the SM. This figure is plotted by counting the number of the samples obtained
in the scan. The left colorful side and the right gray side of each violin are based on 20889 and 8517 samples, respectively, which are marked by the royal blue triangle and the sky blue star in Fig.~\ref{fig1}. The widths of both sides are fixed so that the ratio of
the two widths does not represent the relative sample number. }
\end{figure}

Finally, we concentrate on the particle mass spectra shown in Fig.~\ref{fig3}. After comparing the normalized mass distributions before and after including the LHC restrictions (corresponding to left and right sides of the violin for each particle), one can infer that the LHC search for SUSY affects little the shape of $m_{\tilde{\chi}_{4,5}^0}$ and $m_{\tilde{\chi}_2^\pm}$ distributions, but prefers clearly more massive $\tilde{\chi}_{1}^0$, $\tilde{\chi}_1^\pm$, $\tilde{\chi}_{2}^0$, $\tilde{\chi}_{3}^0$, and $\tilde{\mu}_{L,R}$. The favored mass range for the latter set of particles is $140~{\rm GeV} \lesssim m_{\tilde{\chi}_1^0} \lesssim 600~{\rm GeV}$,  $200~{\rm GeV} \lesssim m_{\tilde{\chi}_2^0}, m_{\tilde{\chi}_1^\pm} \lesssim 700~{\rm GeV}$,  $250~{\rm GeV} \lesssim m_{\tilde{\chi}_3^0} \lesssim 900~{\rm GeV}$,  and $255~{\rm GeV} \lesssim m_{\tilde{\mu}_L} \lesssim 1000~{\rm GeV}$, where the upper bounds come from the requirement to explain the muon g-2 anomaly~\cite{Chakraborti:2020vjp}. We verify that $\tilde{\chi}_{2,3}^0$ and $\tilde{\chi}_1^\pm$ are higgsino-dominated in most cases.  Fig.~\ref{fig3} also indicates $140~{\rm GeV} \lesssim m_{h_s} \lesssim 550~{\rm GeV}$ and $10~{\rm GeV} \lesssim m_{A_s} \lesssim 850~{\rm GeV}$. These mass regions are favored by natural electroweak symmetry breaking. They also
enable the scalars to play a crucial role in the DM annihilations.

\begin{figure}[t]
	\centering
	\includegraphics[width=0.45\textwidth]{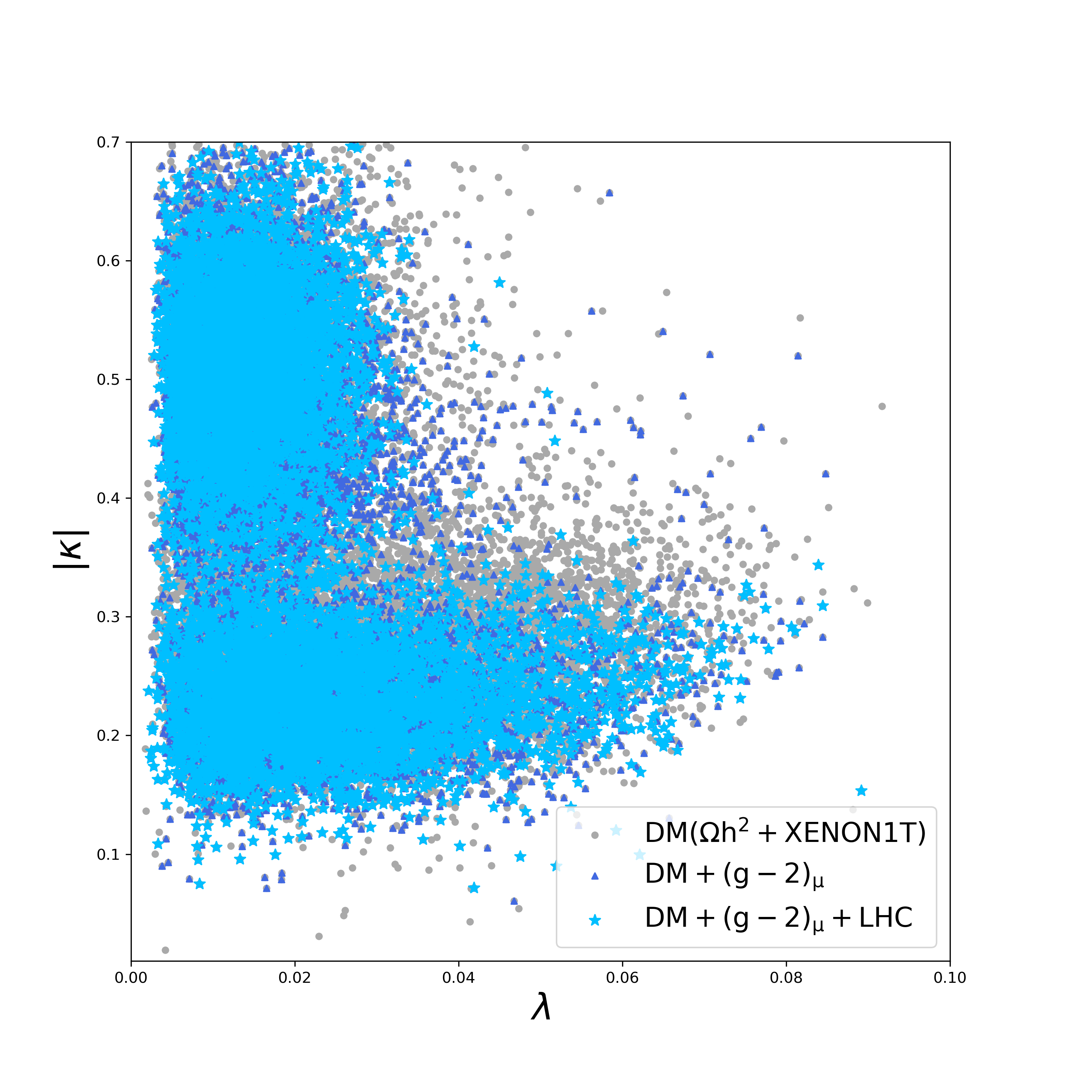}\hspace{-0.3cm}
	\includegraphics[width=0.45\textwidth]{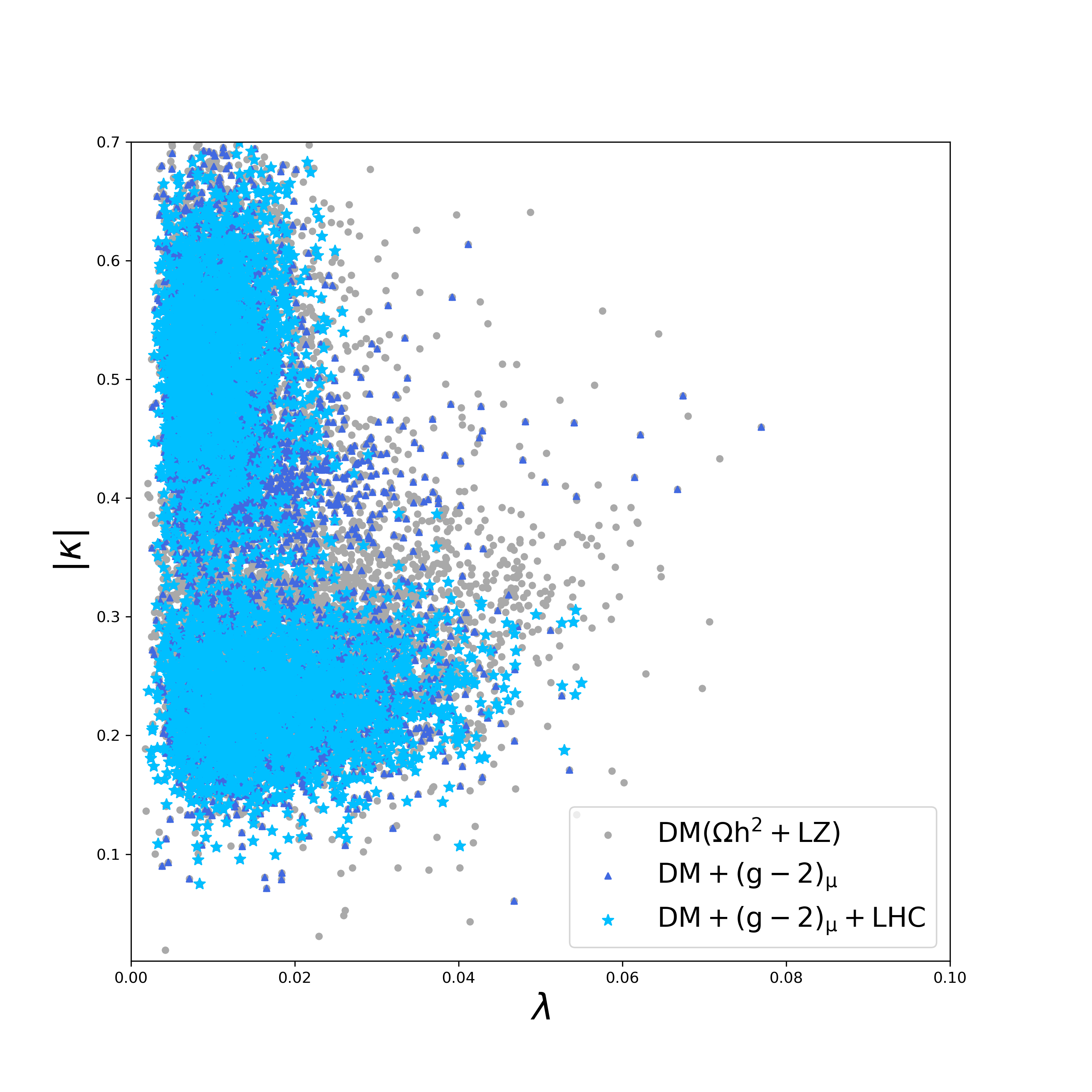}%\hspace{-0.3cm}

\vspace{-0.5cm}

	\caption{\label{fig4} Left panel: similar to Fig.~\ref{fig1}, but the samples are projected onto the $\lambda-|\kappa|$ plane. Right panel: same as the left panel, but it is for the samples satisfying the LZ constraints. }
\end{figure}

\begin{table}[tpb]
	\centering
	\caption{\label{Table5} Same as Table~\ref{Table4}, except that the considered samples satisfy the LZ constraints. }

\hspace{0.1cm}

	\begin{tabular}{l|c|c}
		\hline
		Annihilation Mechanisms& Without LHC Constraints &  With LHC Constraints       \\ \hline
		\multicolumn{1}{l|}{Total Samples}                                                                        &9694             &4166\\
		{$\tilde{\chi}_1^0\tilde{\chi}_1^0 \rightarrow h_{s}h_{s}$}                                                                 &4553              &2207\\
		\multicolumn{1}{l|}{$\tilde{\chi}_1^0\tilde{\chi}_1^0 \rightarrow h_{s}A_{s}$}                                                                 &4545                 &1736\\
		\multicolumn{1}{l|}{$\tilde{\chi}_1^0\tilde{\chi}_1^0 \rightarrow hA_{s}$}
		&4              &0\\
		%\multicolumn{1}{l|}{$A_{s}$-funnel}                                          &26              &7\\
		\multicolumn{1}{l|}{$A_{s}$-funnel}                                          &128              &37\\
		\multicolumn{1}{l|}{$\tilde{\chi}_1^0\tilde{\chi}_1^0 \rightarrow A_{s}A_{s}$}                                                                 &3              &0\\	
		\multicolumn{1}{l|}{$\tilde{\chi}_1^0\tilde{\chi}_1^0 \rightarrow hh_{s}$}
		&1              &0\\
		\multicolumn{1}{l|}{Higgsino co-annihilation}                                                                    &235              &142 \\
		\multicolumn{1}{l|}{Wino co-annihilation}                                                                    &55              &41 \\
		\multicolumn{1}{l|}{$\tilde{\nu}_\mu/\tilde{\mu}_L$ co-annihilation}                                                                    &158              &3 \\
		\multicolumn{1}{l|}{$\tilde{\mu}_R$ co-annihilation}                                                                    &12             &0 \\ \hline
	\end{tabular}
\end{table}

\subsection{\label{numerical study1}Impact of the LZ experiment}

In the last section, the XENON-1T results were used to set upper bounds on the SI and SD cross-sections of the DM-nucleon scattering~\cite{Aprile:2018dbl,Aprile:2019dbj}.
These limitations, however, have been improved significantly by the recent LZ experiment~\cite{LZ:2022ufs}. So in this section, we study the impact of the LZ restrictions on the status of the GNMSSM. Specifically, we utilize the LZ results to further select the samples obtained in the last section, project them onto various panels, and compare the resulting figures with their corresponding ones in the last section.
We conclude that the only difference comes from the distribution of $\lambda$: if the XENON-1T experiment is used to limit the parameter space, $\lambda$ is upper bounded by about 0.09, while if the LZ constraints are adopted, it is less than 0.06\footnote{The distributions of the other parameters are quite similar to those in the last section. In particular, the mass spectra distributions in Fig.~\ref{fig3} are only slightly changed.}. This difference is shown in Fig.~\ref{fig4}, where the samples are projected onto the $\lambda-\kappa$ planes. It arises from the facts that the cross-sections of the DM-nucleon scattering are proportional to $\lambda^4$~\cite{Cao:2021ljw}, and a small $\lambda$ hardly affects the other observables. Due to the difference, the number of the total samples in Fig.~\ref{fig1} is reduced by a factor of about 2. In Table~\ref{Table5}, we list the distribution of the remaining samples by the DM annihilation mechanisms in analogue to Table~\ref{Table4}. This table shows similar features to Table~\ref{Table4}. In particular, both tables show that the LHC restrictions are most efficient in excluding the slepton co-annihilation cases, and least efficient in limiting the electroweakino co-annihilation cases. We explained this phenomenon in~\cite{Cao:2022chy}.

\begin{table}[t]
	\caption{\label{Table6}Detailed information of two benchmark points, P1 and P2, that are consistent with all experimental data. These two points are characterized by a moderately large $\tan \beta$.}
	\vspace{0.3cm}
	\resizebox{1.0\textwidth}{!}{
		\begin{tabular}{llll|llll}
			\hline\hline
			\multicolumn{4}{l|}{\texttt{Benchmark Point P1}}& \multicolumn{4}{l}{\texttt{Benchmark Point P2}}                                                                        \\ \hline
			\multicolumn{1}{l}{$\lambda$}     & \multicolumn{1}{r}{0.018} & \multicolumn{1}{l}{$\text{m}_{\text{h}_s}$}        &\multicolumn{1}{r|}{246.0 GeV}     & \multicolumn{1}{l}{$\lambda$}     & \multicolumn{1}{r}{0.01} & \multicolumn{1}{l}{$\text{m}_{\text{h}_s}$}        &\multicolumn{1}{r}{215.2 GeV}     \\
			\multicolumn{1}{l}{$\kappa$}     & \multicolumn{1}{r}{-0.22} & \multicolumn{1}{l}{$\text{m}_{\text{A}_s}$}        &\multicolumn{1}{r|}{100.0 GeV}     & \multicolumn{1}{l}{$\kappa$}     & \multicolumn{1}{r}{0.48} & \multicolumn{1}{l}{$\text{m}_{\text{A}_s}$}        &\multicolumn{1}{r}{532.6 GeV}     \\
			\multicolumn{1}{l}{$\text{tan}\beta$}      & \multicolumn{1}{r}{29.0} & \multicolumn{1}{l}{$\text{m}_\text{h}$}         &\multicolumn{1}{r|}{125.4 GeV}     & \multicolumn{1}{l}{$\text{tan}\beta$}      & \multicolumn{1}{r}{29.8} & \multicolumn{1}{l}{$\text{m}_\text{h}$}         &\multicolumn{1}{r}{124.8 GeV}     \\
			\multicolumn{1}{l}{$\mu$}        & \multicolumn{1}{r}{330.7 GeV} & \multicolumn{1}{l}{$\text{m}_\text{H}$}         &\multicolumn{1}{r|}{1444.7 GeV}     & \multicolumn{1}{l}{$\mu$}        & \multicolumn{1}{r}{318.9 GeV} & \multicolumn{1}{l}{$\text{m}_\text{H}$}         &\multicolumn{1}{r}{1351.4 GeV}     \\
			\multicolumn{1}{l}{$\mu_{\text{tot}}$} & \multicolumn{1}{r}{336.6 GeV} & \multicolumn{1}{l}{$\text{m}_{\text{A}_\text{H}}$}        &\multicolumn{1}{r|}{1444.7 GeV}     & \multicolumn{1}{l}{$\mu_{\text{tot}}$} & \multicolumn{1}{r}{320.2 GeV} & \multicolumn{1}{l}{$\text{m}_{\text{A}_\text{H}}$}        &\multicolumn{1}{r}{1351.4 GeV}     \\
			\multicolumn{1}{l}{$\text{A}_t$}        & \multicolumn{1}{r}{2544.2 GeV} & \multicolumn{1}{l}{$\text{m}_{\tilde{\chi}_1^0}$}        &\multicolumn{1}{r|}{-271.1 GeV}     & \multicolumn{1}{l}{$\text{A}_t$}        & \multicolumn{1}{r}{2389.2 GeV} & \multicolumn{1}{l}{$\text{m}_{\tilde{\chi}_1^0}$}        &\multicolumn{1}{r}{237.9 GeV}     \\
			\multicolumn{1}{l}{$\text{A}_{\kappa}$}        & \multicolumn{1}{r}{-192.1 GeV} & \multicolumn{1}{l}{$\text{m}_{\tilde{\chi}_2^0}$}        &\multicolumn{1}{r|}{333.7 GeV}     & \multicolumn{1}{l}{$\text{A}_{\kappa}$}        & \multicolumn{1}{r}{266.8 GeV} & \multicolumn{1}{l}{$\text{m}_{\tilde{\chi}_2^0}$}        &\multicolumn{1}{r}{315.9 GeV}     \\
			\multicolumn{1}{l}{$\text{M}_1$}        & \multicolumn{1}{r}{-1251.4 GeV} & \multicolumn{1}{l}{$\text{m}_{\tilde{\chi}_3^0}$}        &\multicolumn{1}{r|}{-347.5 GeV}     & \multicolumn{1}{l}{$\text{M}_1$}        & \multicolumn{1}{r}{1137.0 GeV} & \multicolumn{1}{l}{$\text{m}_{\tilde{\chi}_3^0}$}        &\multicolumn{1}{r}{-332.3 GeV}     \\
			\multicolumn{1}{l}{$\text{M}_2$}        & \multicolumn{1}{r}{565.0 GeV} & \multicolumn{1}{l}{$\text{m}_{\tilde{\chi}_4^0}$}        &\multicolumn{1}{r|}{605.9 GeV}     & \multicolumn{1}{l}{$\text{M}_2$}        & \multicolumn{1}{r}{564.8 GeV} & \multicolumn{1}{l}{$\text{m}_{\tilde{\chi}_4^0}$}        &\multicolumn{1}{r}{604.8 GeV}     \\
			\multicolumn{1}{l}{$\tilde{\text{m}}_{\tilde{\mu}_{\text{L}}}$}        & \multicolumn{1}{r}{630.0 GeV} & \multicolumn{1}{l}{$\text{m}_{\tilde{\chi}_5^0}$}        &\multicolumn{1}{r|}{-1255.9 GeV}     & \multicolumn{1}{l}{$\tilde{\text{m}}_{\tilde{\mu}_{\text{L}}}$}        & \multicolumn{1}{r}{623.6 GeV} & \multicolumn{1}{l}{$\text{m}_{\tilde{\chi}_5^0}$}        &\multicolumn{1}{r}{1141.1 GeV}     \\
			\multicolumn{1}{l}{$\tilde{\text{m}}_{\tilde{\mu}_{\text{R}}}$}        & \multicolumn{1}{r}{997.9 GeV} & \multicolumn{1}{l}{$\text{m}_{\tilde{\chi}_1^{\pm}}$}        &\multicolumn{1}{r|}{336.2 GeV}     & \multicolumn{1}{l}{$\tilde{\text{m}}_{\tilde{\mu}_{\text{R}}}$}        & \multicolumn{1}{r}{868.0 GeV} & \multicolumn{1}{l}{$\text{m}_{\tilde{\chi}_1^{\pm}}$}      &\multicolumn{1}{r}{320.2 GeV}     \\
			\multicolumn{1}{l}{$a_{\mu}^{\text{SUSY}}$}       & \multicolumn{1}{r}{$1.34 \times 10^{-9}$} & \multicolumn{1}{l}{$\text{m}_{\tilde{\chi}_2^{\pm}}$}        &\multicolumn{1}{r|}{606.2 GeV}     & \multicolumn{1}{l}{$a_{\mu}^{\text{SUSY}}$}       & \multicolumn{1}{r}{$1.46 \times 10^{-9}$} & \multicolumn{1}{l}{$\text{m}_{\tilde{\chi}_2^{\pm}}$}       &\multicolumn{1}{r}{605.3 GeV}     \\
			\multicolumn{1}{l}{${\Omega h}^2$}     & \multicolumn{1}{r}{0.12} & \multicolumn{1}{l}{$\text{m}_{\tilde{\mu}_{\text{L}}}$}      &\multicolumn{1}{r|}{635.9 GeV}     & \multicolumn{1}{l}{${\Omega h}^2$}     & \multicolumn{1}{r}{0.13} & \multicolumn{1}{l}{$\text{m}_{\tilde{\mu}_{\text{L}}}$}    &\multicolumn{1}{r}{631.6 GeV}     \\
			\multicolumn{1}{l}{$\sigma_{p}^{\text{SI}}$}   & \multicolumn{1}{r}{$1.83 \times 10^{-47}\text{cm}^{2}$} & \multicolumn{1}{l}{$\text{m}_{\tilde{\mu}_{\text{R}}}$}      &\multicolumn{1}{r|}{1002.7 GeV}     & \multicolumn{1}{l}{$\sigma_{p}^{\text{SI}}$}   &\multicolumn{1}{r}{$1.71 \times 10^{-47}\text{cm}^{2}$}& \multicolumn{1}{l}{$\text{m}_{\tilde{\mu}_{\text{R}}}$}    &\multicolumn{1}{r}{873.2 GeV}     \\
			\multicolumn{1}{l}{$\sigma_{n}^{\text{SD}}$}   & \multicolumn{1}{r}{$1.44 \times 10^{-45}\text{cm}^2$} & \multicolumn{1}{l}{$\text{m}_{\tilde{\nu}_{\mu}}$}    &\multicolumn{1}{r|}{630.7 GeV}     & \multicolumn{1}{l}{$\sigma_{n}^{\text{SD}}$}   & \multicolumn{1}{r}{$1.07 \times 10^{-46}\text{cm}^2$} & \multicolumn{1}{l}{$\text{m}_{\tilde{\nu}_{\mu}}$}    &\multicolumn{1}{r}{626.3 GeV}     \\ \hline
			\multicolumn{2}{l}{$\text{N}_{11},\text{N}_{12},\text{N}_{13},\text{N}_{14},\text{N}_{15}$}           & \multicolumn{2}{l|}{0.0011, 0.0021, 0.0186, 0.0237, -0.9995}                 & \multicolumn{2}{l}{$\text{N}_{11},\text{N}_{12},\text{N}_{13},\text{N}_{14},\text{N}_{15}$}           & \multicolumn{2}{l}{0.0006, -0.0028, 0.0093, -0.0121, 0.9999}                 \\
			\multicolumn{2}{l}{$\text{N}_{21},\text{N}_{22},\text{N}_{23},\text{N}_{24},\text{N}_{25}$}           & \multicolumn{2}{l|}{-0.019, -0.216, 0.701, -0.679, -0.004}                 & \multicolumn{2}{l}{$\text{N}_{21},\text{N}_{22},\text{N}_{23},\text{N}_{24},\text{N}_{25}$}           & \multicolumn{2}{l}{0.037, -0.202, 0.705, -0.679, -0.015}                 \\
			\multicolumn{2}{l}{$\text{N}_{31},\text{N}_{32},\text{N}_{33},\text{N}_{34},\text{N}_{35}$}           & \multicolumn{2}{l|}{0.033, 0.058, 0.703, 0.707, 0.030}                 & \multicolumn{2}{l}{$\text{N}_{31},\text{N}_{32},\text{N}_{33},\text{N}_{34},\text{N}_{35}$}           & \multicolumn{2}{l}{-0.020, 0.059, 0.702, 0.710, 0.002}                 \\
			\multicolumn{2}{l}{$\text{N}_{41},\text{N}_{42},\text{N}_{43},\text{N}_{44},\text{N}_{45}$}           & \multicolumn{2}{l|}{-0.005, 0.975, 0.114, -0.193, -0.0004}                 & \multicolumn{2}{l}{$\text{N}_{41},\text{N}_{42},\text{N}_{43},\text{N}_{44},\text{N}_{45}$}           & \multicolumn{2}{l}{0.015,  -0.006, 0.103, -0.183, -0.0005}                 \\
			\multicolumn{2}{l}{$\text{N}_{51},\text{N}_{52},\text{N}_{53},\text{N}_{54},\text{N}_{55}$}           & \multicolumn{2}{l|}{0.999, -0.002, -0.009, -0.037, 2.54E-05}                 & \multicolumn{2}{l}{$\text{N}_{51},\text{N}_{52},\text{N}_{53},\text{N}_{54},\text{N}_{55}$}           & \multicolumn{2}{l}{0.999, -0.006, -0.014, 0.043, -8.42E-06}                 \\
			\hline
			\multicolumn{2}{l}{Annihilations}                     & \multicolumn{2}{l|}{Fractions[\%]}        & \multicolumn{2}{l}{Annihilations}                     & \multicolumn{2}{l}{Fractions[\%]}        \\
			\multicolumn{2}{l}{$\tilde{\chi}_1^0\tilde{\chi}_1^0 \to h_{s} \text{A}_{s}/h_{s}h_{s}$}                  & \multicolumn{2}{l|}{96.0/3.1}                 & \multicolumn{2}{l}{$\tilde{\chi}_1^0\tilde{\chi}_1^0 \to h_{s} \text{h}_{s}$}                  & \multicolumn{2}{l}{99.6}                 \\
			\hline
			\multicolumn{2}{l}{Decays}                            & \multicolumn{2}{l|}{Branching ratios[\%]} & \multicolumn{2}{l}{Decays}                            & \multicolumn{2}{l}{Branching ratios[\%]} \\
			\multicolumn{2}{l}{$\tilde{\chi}_2^0 \to \tilde{\chi}_1^0Z^{\star}(\to f\bar{f})$}& \multicolumn{2}{l|}{100}
			&\multicolumn{2}{l}{$\tilde{\chi}_2^0 \to \tilde{\chi}_1^0Z^{\star}(\to f\bar{f})$}& \multicolumn{2}{l}{100}                 \\
			\multicolumn{2}{l}{$\tilde{\chi}_3^0 \to \tilde{\chi}_2^0Z^{\star}/\tilde{\chi}_1^0 Z^{\star}/\tilde{\chi}_1^{\pm}W^{\mp\star}$}& \multicolumn{2}{l|}{39.5/39.3/21.2}                 & \multicolumn{2}{l}{$\tilde{\chi}_3^0 \to \tilde{\chi}_1^0Z/\tilde{\chi}_2^0 Z^{\star}/\tilde{\chi}_1^{\pm}W^{\mp\star}$}                           & \multicolumn{2}{l}{77.4/14.4/8.2}                 \\
			\multicolumn{2}{l}{$\tilde{\chi}_4^0 \to \tilde{\chi}_1^{\pm}W^{\mp}/\tilde{\chi}_3^0 Z/\tilde{\chi}_2^0 h/\tilde{\chi}_2^0 Z/\tilde{\chi}_3^0 h$}& \multicolumn{2}{l|}{55.4/22.0/19.5/1.9/1.1}                 & \multicolumn{2}{l}{$\tilde{\chi}_4^0 \to \tilde{\chi}_1^{\pm}W^{\mp}/\tilde{\chi}_3^0 Z/\tilde{\chi}_2^0 h/\tilde{\chi}_2^0 Z/\tilde{\chi}_3^0 h$}                           & \multicolumn{2}{l}{56.3/21.4/18.8/2.1/1.3}                 \\
			\multicolumn{2}{l}{$\tilde{\chi}_5^0 \to \tilde{\chi}_1^{\pm}W^{\mp}/\tilde{\nu}_{\mu}\nu_{\mu}/\tilde{\mu}^{\pm}_{\text{L}}\mu^{\mp}/\tilde{\mu}^{\pm}_{\text{R}}\mu^{\mp}/\tilde{\chi}_3^0 h/\tilde{\chi}_4^0 h/\tilde{\chi}_3^0 Z/\tilde{\chi}_2 h/\tilde{\chi}_4^0 Z$}                           & \multicolumn{2}{l|}{25.1/16.1/15.7/15.0/9.9/9.0/3.6/3.6/3.3}                 & \multicolumn{2}{l}{$\tilde{\chi}_5^0 \to \tilde{\chi}_1^{\pm}W^{\mp}/\tilde{\mu}^{\pm}_{\text{R}}\mu^{\mp}/\tilde{\nu}_{\mu}\nu_{\mu}/\tilde{\mu}^{\pm}_{\text{L}}\mu^{\mp}/\tilde{\chi}_3^0 Z/\tilde{\chi}_2^0 h/\tilde{\chi}_2^{\pm}W^{\mp}/\tilde{\chi}_2^0 Z/\tilde{\chi}_2^0 Z$}                           & \multicolumn{2}{l}{22.9/18.9/13.7/12.9/10.7/9.7/3.2/2.9/2.7}                 \\
			\multicolumn{2}{l}{$\tilde{\chi}_1^{\pm} \to \tilde{\chi}_1^0W^{\pm\star}$}                           & \multicolumn{2}{l|}{100}                 & \multicolumn{2}{l}{$\tilde{\chi}_1^{\pm} \to \tilde{\chi}_1^0W^{\pm}$}                           & \multicolumn{2}{l}{100}                 \\
			\multicolumn{2}{l}{$\tilde{\chi}_2^{\pm} \to \tilde{\chi}_2^0W^{\pm}/\tilde{\chi}_3^0W^{\pm}/\tilde{\chi}_1^{\pm}Z/\tilde{\chi}_1^{\pm}h$}                     & \multicolumn{2}{l|}{26.9/25.8/25.6/21.7}                 & \multicolumn{2}{l}{$\tilde{\chi}_2^{\pm} \to \tilde{\chi}_2^0W^{\pm}/\tilde{\chi}_3^0W^{\pm}/\tilde{\chi}_1^{\pm}Z/\tilde{\chi}_1^{\pm}h$}                     & \multicolumn{2}{l}{26.3/25.8/25.7/22.1}                 \\
			\multicolumn{2}{l}{$\tilde{\mu}_{\text{L}}^{\pm} \to \tilde{\chi}_2^0\mu^{\pm}/\tilde{\chi}_1^{\pm}\nu_{\mu}/\tilde{\chi}_2^{\pm}\nu_{\mu}/\tilde{\chi}_4^{0}\mu^{\pm}/\tilde{\chi}_3^{0}\mu^{\pm}$}                              & \multicolumn{2}{l|}{33.4/33.1/19.7/9.8/4.0}                 & \multicolumn{2}{l}{$\tilde{\mu}_{\text{L}}^{\pm} \to \tilde{\chi}_1^{\pm}\nu_{\mu}/\tilde{\chi}_2^0\mu^{\pm}/\tilde{\chi}_2^{\pm}\nu_{\mu}/\tilde{\chi}_4^0\mu^{\pm}/\tilde{\chi}_3^0\mu^{\pm}$}                              & \multicolumn{2}{l}{37.7/29.3/20.1/10.4/2.5}                 \\
			\multicolumn{2}{l}{$\tilde{\mu}_{\text{R}}^{\pm} \to \tilde{\chi}_3^0\mu^{\pm}/\tilde{\chi}_1^{\pm}\nu_{\mu}/\tilde{\chi}_2^0\mu^{\pm}/\tilde{\nu}_{\mu}W^{\pm}/\tilde{\mu}_{\text{L}}^{\pm}h/\tilde{\mu}_{\text{L}}^{\pm}Z$}                              & \multicolumn{2}{l|}{40.3/29.1/23.6/2.5/1.2/1.2}                 & \multicolumn{2}{l}{$\tilde{\mu}_{\text{R}}^{\pm} \to \tilde{\chi}_2^0\mu^{\pm}/\tilde{\chi}_1^{\pm}\nu_{\mu}/\tilde{\chi}_3^0\mu^{\pm}/\tilde{\nu}_{\mu}W^{\pm}/\tilde{\chi}_4^0\mu^{\pm}/\tilde{\mu}_{\text{L}}^{\pm}h$} &\multicolumn{2}{l}{43.5/26.9/22.5/2.2/2.1/1.0}\\
			\multicolumn{2}{l}{$\tilde{\nu}_{\mu} \to \tilde{\chi}_1^{\pm}\mu^{\mp}/ \tilde{\chi}_2^{0}\nu_{\mu}/\tilde{\chi}_2^{\pm}\mu^{\mp}/\tilde{\chi}_4^{0}\nu_{\mu}$}
			& \multicolumn{2}{l|}{66.7/18.7/9.1/4.9}                 & \multicolumn{2}{l}{$\tilde{\nu}_{\mu} \to \tilde{\chi}_1^{\pm}\mu^{\mp}/\tilde{\chi}_2^0\nu/\tilde{\chi}_2^{\pm}\mu^{\mp}/\tilde{\chi}_4^{0}\nu_{\mu}/\tilde{\chi}_3^{0}\nu_{\mu}$}                             & \multicolumn{2}{l}{64.3/23.2/6.8/3.6/2.2}                 \\
			\hline
			\multicolumn{2}{l}{R value}                           & \multicolumn{2}{l|}{0.48, ATLAS-2106-01676}                 & \multicolumn{2}{l}{R value}                           & \multicolumn{2}{l}{0.51, CMS-SUS-16-039}                 \\ \hline\hline
	\end{tabular}}
\end{table}

At this stage, we emphasize several points about the LHC search for SUSY:
\begin{itemize}
\item There are at least two reasons why the LHC restrictions are so strong in excluding the samples. One is that we include many experimental analyses in this study, and each of them usually defines several signal regions. As a result, some of these analyses are complementary in surveying/restricting SUSY signals even when the same set of experimental data are studied. The other is explaining the muon g-2 anomaly requires more than one moderately light sparticle, which is particularly so if $\tan \beta$ is small. Consequently, the signals of different sparticles superimpose to strengthen their detection at the LHC.

\item Throughout this study, both the theoretical uncertainties incurred by the simulations and the experimental (systematic and statistical) uncertainties are not considered. Although these effects can relax the LHC restrictions, it is expected that much more stringent constraints on the GNMSSM will be obtained in the near future, given the advent of the high-luminosity LHC.

\item In some unification theories, $\tilde{\tau}$ may be the next lightest supersymmetric particle (NLSP) due to its larger Yukawa coupling than the first-two-generation sleptons. In this case, heavy sparticles may decay into $\tilde{\tau}$ to change the production rate of the $e/\mu$ signals. Consequently, the LHC constraints can be relaxed~\cite{Hagiwara:2017lse}. We will discuss such a possibility in future works.
\end{itemize}

\begin{table}[t]
	\caption{\label{Table7}Detailed information of another two benchmark points, P3 and P4, satisfying all experimental constraints. These two points are characterized by a large $\tan \beta$.}
	\vspace{0.3cm}
	\resizebox{1.0\textwidth}{!}{
		\begin{tabular}{llll|llll}
			\hline\hline
			\multicolumn{4}{l|}{\texttt{Benchmark Point P3}}& \multicolumn{4}{l}{\texttt{Benchmark Point P4}}                                                                        \\ \hline
			\multicolumn{1}{l}{$\lambda$}     & \multicolumn{1}{r}{0.012} & \multicolumn{1}{l}{$\text{m}_{\text{h}_s}$}        &\multicolumn{1}{r|}{248.6 GeV}     & \multicolumn{1}{l}{$\lambda$}     & \multicolumn{1}{r}{0.005} & \multicolumn{1}{l}{$\text{m}_{\text{h}_s}$}        &\multicolumn{1}{r}{216.5 GeV}     \\
			\multicolumn{1}{l}{$\kappa$}     & \multicolumn{1}{r}{0.25} & \multicolumn{1}{l}{$\text{m}_{\text{A}_s}$}        &\multicolumn{1}{r|}{162.2 GeV}     & \multicolumn{1}{l}{$\kappa$}     & \multicolumn{1}{r}{-0.44} & \multicolumn{1}{l}{$\text{m}_{\text{A}_s}$}        &\multicolumn{1}{r}{314.8 GeV}     \\
			\multicolumn{1}{l}{$\text{tan}\beta$}      & \multicolumn{1}{r}{51.3} & \multicolumn{1}{l}{$\text{m}_\text{h}$}         &\multicolumn{1}{r|}{125.2 GeV}     & \multicolumn{1}{l}{$\text{tan}\beta$}      & \multicolumn{1}{r}{55.4} & \multicolumn{1}{l}{$\text{m}_\text{h}$}         &\multicolumn{1}{r}{125.5 GeV}     \\
			\multicolumn{1}{l}{$\mu$}        & \multicolumn{1}{r}{384.4 GeV} & \multicolumn{1}{l}{$\text{m}_\text{H}$}         &\multicolumn{1}{r|}{2235.7 GeV}     & \multicolumn{1}{l}{$\mu$}        & \multicolumn{1}{r}{343.1 GeV} & \multicolumn{1}{l}{$\text{m}_\text{H}$}         &\multicolumn{1}{r}{2077.0 GeV}     \\
			\multicolumn{1}{l}{$\mu_{\text{tot}}$} & \multicolumn{1}{r}{393.1 GeV} & \multicolumn{1}{l}{$\text{m}_{\text{A}_\text{H}}$}        &\multicolumn{1}{r|}{2235.7 GeV}     & \multicolumn{1}{l}{$\mu_{\text{tot}}$} & \multicolumn{1}{r}{344.4 GeV} & \multicolumn{1}{l}{$\text{m}_{\text{A}_\text{H}}$}        &\multicolumn{1}{r}{2077.0 GeV}     \\
			\multicolumn{1}{l}{$\text{A}_t$}        & \multicolumn{1}{r}{2540.9 GeV} & \multicolumn{1}{l}{$\text{m}_{\tilde{\chi}_1^0}$}        &\multicolumn{1}{r|}{271.8 GeV}     & \multicolumn{1}{l}{$\text{A}_t$}        & \multicolumn{1}{r}{2533.8 GeV} & \multicolumn{1}{l}{$\text{m}_{\tilde{\chi}_1^0}$}        &\multicolumn{1}{r}{-237.5 GeV}     \\
			\multicolumn{1}{l}{$\text{A}_{\kappa}$}        & \multicolumn{1}{r}{-112.1 GeV} & \multicolumn{1}{l}{$\text{m}_{\tilde{\chi}_2^0}$}        &\multicolumn{1}{r|}{402.4 GeV}     & \multicolumn{1}{l}{$\text{A}_{\kappa}$}        & \multicolumn{1}{r}{47.0 GeV} & \multicolumn{1}{l}{$\text{m}_{\tilde{\chi}_2^0}$}        &\multicolumn{1}{r}{346.8 GeV}     \\
			\multicolumn{1}{l}{$\text{M}_1$}        & \multicolumn{1}{r}{-854.6 GeV} & \multicolumn{1}{l}{$\text{m}_{\tilde{\chi}_3^0}$}        &\multicolumn{1}{r|}{-404.3 GeV}     & \multicolumn{1}{l}{$\text{M}_1$}        & \multicolumn{1}{r}{796.9 GeV} & \multicolumn{1}{l}{$\text{m}_{\tilde{\chi}_3^0}$}        &\multicolumn{1}{r}{-358.0 GeV}     \\
			\multicolumn{1}{l}{$\text{M}_2$}        & \multicolumn{1}{r}{1267.9 GeV} & \multicolumn{1}{l}{$\text{m}_{\tilde{\chi}_4^0}$}        &\multicolumn{1}{r|}{-858.8 GeV}     & \multicolumn{1}{l}{$\text{M}_2$}        & \multicolumn{1}{r}{826.3 GeV} & \multicolumn{1}{l}{$\text{m}_{\tilde{\chi}_4^0}$}        &\multicolumn{1}{r}{801.0 GeV}     \\
			\multicolumn{1}{l}{$\tilde{\text{m}}_{\tilde{\mu}_{\text{L}}}$}        & \multicolumn{1}{r}{390.0 GeV} & \multicolumn{1}{l}{$\text{m}_{\tilde{\chi}_5^0}$}        &\multicolumn{1}{r|}{1303.3 GeV}     & \multicolumn{1}{l}{$\tilde{\text{m}}_{\tilde{\mu}_{\text{L}}}$}        & \multicolumn{1}{r}{580.1 GeV} & \multicolumn{1}{l}{$\text{m}_{\tilde{\chi}_5^0}$}        &\multicolumn{1}{r}{864.4 GeV}     \\
			\multicolumn{1}{l}{$\tilde{\text{m}}_{\tilde{\mu}_{\text{R}}}$}        & \multicolumn{1}{r}{569.7 GeV} & \multicolumn{1}{l}{$\text{m}_{\tilde{\chi}_1^{\pm}}$}        &\multicolumn{1}{r|}{403.5 GeV}     & \multicolumn{1}{l}{$\tilde{\text{m}}_{\tilde{\mu}_{\text{R}}}$}        & \multicolumn{1}{r}{715.4 GeV} & \multicolumn{1}{l}{$\text{m}_{\tilde{\chi}_1^{\pm}}$}      &\multicolumn{1}{r}{351.5 GeV}     \\
			\multicolumn{1}{l}{$a_{\mu}^{\text{SUSY}}$}       & \multicolumn{1}{r}{$1.42 \times 10^{-9}$} & \multicolumn{1}{l}{$\text{m}_{\tilde{\chi}_2^{\pm}}$}        &\multicolumn{1}{r|}{1303.5 GeV}     & \multicolumn{1}{l}{$a_{\mu}^{\text{SUSY}}$}       & \multicolumn{1}{r}{$2.15 \times 10^{-9}$} & \multicolumn{1}{l}{$\text{m}_{\tilde{\chi}_2^{\pm}}$}       &\multicolumn{1}{r}{864.2 GeV}     \\
			\multicolumn{1}{l}{${\Omega h}^2$}     & \multicolumn{1}{r}{0.11} & \multicolumn{1}{l}{$\text{m}_{\tilde{\mu}_{\text{L}}}$}      &\multicolumn{1}{r|}{405.4 GeV}     & \multicolumn{1}{l}{${\Omega h}^2$}     & \multicolumn{1}{r}{0.13} & \multicolumn{1}{l}{$\text{m}_{\tilde{\mu}_{\text{L}}}$}    &\multicolumn{1}{r}{591.8 GeV}     \\
			\multicolumn{1}{l}{$\sigma_{p}^{\text{SI}}$}   & \multicolumn{1}{r}{$1.67 \times 10^{-47}\text{cm}^{2}$} & \multicolumn{1}{l}{$\text{m}_{\tilde{\mu}_{\text{R}}}$}      &\multicolumn{1}{r|}{575.1 GeV}     & \multicolumn{1}{l}{$\sigma_{p}^{\text{SI}}$}   &\multicolumn{1}{r}{$1.36 \times 10^{-47}\text{cm}^{2}$}& \multicolumn{1}{l}{$\text{m}_{\tilde{\mu}_{\text{R}}}$}    &\multicolumn{1}{r}{720.2 GeV}     \\
			\multicolumn{1}{l}{$\sigma_{n}^{\text{SD}}$}   & \multicolumn{1}{r}{$7.52 \times 10^{-47}\text{cm}^2$} & \multicolumn{1}{l}{$\text{m}_{\tilde{\nu}_{\mu}}$}    &\multicolumn{1}{r|}{397.4 GeV}     & \multicolumn{1}{l}{$\sigma_{n}^{\text{SD}}$}   & \multicolumn{1}{r}{$4.68 \times 10^{-48}\text{cm}^2$} & \multicolumn{1}{l}{$\text{m}_{\tilde{\nu}_{\mu}}$}    &\multicolumn{1}{r}{586.2 GeV}     \\ \hline
			\multicolumn{2}{l}{$\text{N}_{11},\text{N}_{12},\text{N}_{13},\text{N}_{14},\text{N}_{15}$}           & \multicolumn{2}{l|}{-0.0004, -0.0007, 0.0064, -0.0095, 0.9999}                 & \multicolumn{2}{l}{$\text{N}_{11},\text{N}_{12},\text{N}_{13},\text{N}_{14},\text{N}_{15}$}           & \multicolumn{2}{l}{-0.0002, 0.0003, 0.0031, 0.0047, -0.9999}                 \\
			\multicolumn{2}{l}{$\text{N}_{21},\text{N}_{22},\text{N}_{23},\text{N}_{24},\text{N}_{25}$}           & \multicolumn{2}{l|}{0.025, 0.063, -0.708, 0.703, 0.011}                 & \multicolumn{2}{l}{$\text{N}_{21},\text{N}_{22},\text{N}_{23},\text{N}_{24},\text{N}_{25}$}           & \multicolumn{2}{l}{-0.068, -0.109, 0.709, -0.693, -0.001}                 \\
			\multicolumn{2}{l}{$\text{N}_{31},\text{N}_{32},\text{N}_{33},\text{N}_{34},\text{N}_{35}$}           & \multicolumn{2}{l|}{-0.067, -0.032, -0.706, -0.705, -0.002}                 & \multicolumn{2}{l}{$\text{N}_{31},\text{N}_{32},\text{N}_{33},\text{N}_{34},\text{N}_{35}$}           & \multicolumn{2}{l}{-0.026, 0.045, 0.703, 0.709, 0.005}                 \\
			\multicolumn{2}{l}{$\text{N}_{41},\text{N}_{42},\text{N}_{43},\text{N}_{44},\text{N}_{45}$}           & \multicolumn{2}{l|}{0.997, -0.003, -0.030, -0.065, -1.33E-05}                 & \multicolumn{2}{l}{$\text{N}_{41},\text{N}_{42},\text{N}_{43},\text{N}_{44},\text{N}_{45}$}           & \multicolumn{2}{l}{-0.993,  -0.099, 0.026, -0.056, -3.93E-06}                 \\
			\multicolumn{2}{l}{$\text{N}_{51},\text{N}_{52},\text{N}_{53},\text{N}_{54},\text{N}_{55}$}           & \multicolumn{2}{l|}{0.001, -0.998, -0.022, 0.067, 4.3E-06}                 & \multicolumn{2}{l}{$\text{N}_{51},\text{N}_{52},\text{N}_{53},\text{N}_{54},\text{N}_{55}$}           & \multicolumn{2}{l}{0.091, -0.988, -0.049, 0.115, 3.82E-05}                 \\
			\hline
			\multicolumn{2}{l}{Annihilations}                     & \multicolumn{2}{l|}{Fractions[\%]}        & \multicolumn{2}{l}{Annihilations}                     & \multicolumn{2}{l}{Fractions[\%]}        \\
			\multicolumn{2}{l}{$\tilde{\chi}_1^0\tilde{\chi}_1^0 \to h_{s} \text{A}_{s}/h_{s}h_{s}/\text{A}_{s}\text{A}_{s}$}                  & \multicolumn{2}{l|}{91.8/6.4/1.7}                 & \multicolumn{2}{l}{$\tilde{\chi}_1^0\tilde{\chi}_1^0 \to h_{s} \text{h}_{s}/h A_{s}/h_{s}A_{s}$}                  & \multicolumn{2}{l}{96.8/2.7/1.7/1.4}                 \\
			\hline
			\multicolumn{2}{l}{Decays}                            & \multicolumn{2}{l|}{Branching ratios[\%]} & \multicolumn{2}{l}{Decays}                            & \multicolumn{2}{l}{Branching ratios[\%]} \\
			\multicolumn{2}{l}{$\tilde{\chi}_2^0 \to \tilde{\chi}_1^0h/\tilde{\chi}_1^0Z/\tilde{\nu}_{\mu}\nu_{\mu}$}& \multicolumn{2}{l|}{83.9/14.5/1.6}
			&\multicolumn{2}{l}{$\tilde{\chi}_2^0 \to \tilde{\chi}_1^0Z$}& \multicolumn{2}{l}{100}                 \\
			\multicolumn{2}{l}{$\tilde{\chi}_3^0 \to \tilde{\chi}_1^0Z$}& \multicolumn{2}{l|}{99.6}                 & \multicolumn{2}{l}{$\tilde{\chi}_3^0 \to \tilde{\chi}_1^0Z/\tilde{\chi}_2^0 Z^{\star}$}                           & \multicolumn{2}{l}{96.3/3.1}                 \\
			\multicolumn{2}{l}{$\tilde{\chi}_4^0 \to \tilde{\mu}^{\pm}_{\text{R}}\mu^{\mp}/\tilde{\chi}_1^{\pm}W^{\mp}/\tilde{\nu}_{\mu}\nu_{\mu}/\tilde{\mu}^{\pm}_{\text{L}}\mu^{\mp}/\tilde{\chi}_3^0 h/\tilde{\chi}_2^0 Z/\tilde{\chi}_3^0 Z/\tilde{\chi}_2^0 h$}& \multicolumn{2}{l|}{29.2/20.6/15.0/14.4/9.2/8.8/1.4/1.2}                 & \multicolumn{2}{l}{$\tilde{\chi}_4^0 \to \tilde{\chi}_1^{\pm}W^{\mp}/\tilde{\mu}^{\pm}_{\text{L}}\mu^{\mp}/\tilde{\chi}_3^0 Z/\tilde{\chi}_2^0 h/\tilde{\mu}^{\pm}_{\text{R}}\mu^{\mp}/\tilde{\nu}_{\mu}\nu_{\mu}/\tilde{\chi}_2^0 Z/\tilde{\chi}_3^0 h$}                           & \multicolumn{2}{l}{50.2/12.1/11.1/10.7/6.2/6.2/1.6/1.5}                 \\
			\multicolumn{2}{l}{$\tilde{\chi}_5^0 \to \tilde{\chi}_1^{\pm}W^{\mp}/\tilde{\nu}_{\mu}\nu_{\mu}/\tilde{\mu}^{\pm}_{\text{L}}\mu^{\mp}/\tilde{\chi}_2^0 h/\tilde{\chi}_3^0 Z/\tilde{\chi}_2 Z/\tilde{\chi}_3^0 h$}                           & \multicolumn{2}{l|}{27.3/22.7/22.4/10.7/10.7/2.8/2.7}                 & \multicolumn{2}{l}{$\tilde{\chi}_5^0 \to \tilde{\chi}_1^{\pm}W^{\mp}/\tilde{\chi}_3^0 Z/\tilde{\chi}_2^0 h/\tilde{\nu}_{\mu}\nu_{\mu}/\tilde{\mu}^{\pm}_{\text{L}}\mu^{\mp}/\tilde{\chi}_3^0 Z/\tilde{\chi}_2^0 Z$}                           & \multicolumn{2}{l}{35.8/17.9/16.7/13.1/10.2/3.1/2.7}                 \\
			\multicolumn{2}{l}{$\tilde{\chi}_1^{\pm} \to \tilde{\chi}_1^0W^{\pm}/\tilde{\nu}_{\mu}\mu^{\pm}$}                           & \multicolumn{2}{l|}{93.3/6.7}                 & \multicolumn{2}{l}{$\tilde{\chi}_1^{\pm} \to \tilde{\chi}_1^0W^{\pm}$}                           & \multicolumn{2}{l}{100}                 \\
			\multicolumn{2}{l}{$\tilde{\chi}_2^{\pm} \to \tilde{\nu}_{\mu}\mu^{\pm}/\tilde{\mu}^{\pm}_{\text{L}}\nu_{\mu}/\tilde{\chi}_3^0W^{\pm}/\tilde{\chi}_2^0W^{\pm}/\tilde{\chi}_1^{\pm}Z/\tilde{\chi}_1^{\pm}h$}                     & \multicolumn{2}{l|}{22.3/22.3/13.6/13.6/13.5/13.5}                 & \multicolumn{2}{l}{$\tilde{\chi}_2^{\pm} \to \tilde{\chi}_3^0W^{\pm}/\tilde{\chi}_1^{\pm}Z/\tilde{\chi}_2^0W^{\pm}/\tilde{\chi}_1^{\pm}h/\tilde{\nu}_{\mu}\mu^{\pm}/\tilde{\mu}^{\pm}_{\text{L}}\nu_{\mu}$}                     & \multicolumn{2}{l}{19.5/19.3/19.0/18.6/11.7/11.5}                 \\
			\multicolumn{2}{l}{$\tilde{\mu}_{\text{L}}^{\pm} \to \tilde{\chi}_2^0\mu^{\pm}/\tilde{\chi}_1^{0}\mu^{\pm}/\tilde{\chi}_1^{\pm}\nu_{\mu}$}                              & \multicolumn{2}{l|}{71.0/13.5/6.6}                 & \multicolumn{2}{l}{$\tilde{\mu}_{\text{L}}^{\pm} \to \tilde{\chi}_1^{\pm}\nu_{\mu}/\tilde{\chi}_2^0\mu^{\pm}/\tilde{\chi}_2^{\pm}\nu_{\mu}/\tilde{\chi}_3^0\mu^{\pm}$}                              & \multicolumn{2}{l}{41.3/40.5/18.2}                 \\
			\multicolumn{2}{l}{$\tilde{\mu}_{\text{R}}^{\pm} \to \tilde{\chi}_3^0\mu^{\pm}/\tilde{\chi}_1^{\pm}\nu_{\mu}/\tilde{\nu}_{\mu}W^{\pm}/\tilde{\chi}_2^0\mu^{\pm}/\tilde{\mu}_{\text{L}}^{\pm}h/\tilde{\mu}_{\text{L}}^{\pm}Z$}                              & \multicolumn{2}{l|}{33.2/22.7/16.5/14.5/7.0/6.1}                 & \multicolumn{2}{l}{$\tilde{\mu}_{\text{R}}^{\pm} \to \tilde{\chi}_2^0\mu^{\pm}/\tilde{\chi}_1^{\pm}\nu_{\mu}/\tilde{\chi}_3^0\mu^{\pm}$} &\multicolumn{2}{l}{45.3/30.9/20.2}\\
			\multicolumn{2}{l}{$\tilde{\nu}_{\mu} \to \tilde{\chi}_1^{0}\nu_{\mu}$}
			& \multicolumn{2}{l|}{100}                 & \multicolumn{2}{l}{$\tilde{\nu}_{\mu} \to \tilde{\chi}_1^{\pm}\mu^{\mp}/\tilde{\chi}_2^0\nu_{\mu}/\tilde{\chi}_3^{0}\nu_{\mu}$}                             & \multicolumn{2}{l}{67.9/27.8/4.3}                 \\
			\hline
			\multicolumn{2}{l}{R value}                           & \multicolumn{2}{l|}{0.18, CMS-SUS-16-039}                 & \multicolumn{2}{l}{R value}                           & \multicolumn{2}{l}{0.62, ATLAS-2106-01676}                 \\ \hline\hline
	\end{tabular}}
\end{table}

\subsection{Benchmark points}

In this section, we provide four parameter points in Tables~\ref{Table6} and~\ref{Table7} to improve understanding the intrinsic physics. These points satisfy all the experimental constraints and exhibit the following common characteristics:
\begin{itemize}
\item The higgsino mass $\mu_{tot}$ is less than $400~{\rm GeV}$ so that the $Z$-boson mass is naturally predicted. As introduced in Sec.~\ref{Introduction}, the MSSM does not have this property.
\item The singlino-dominated DM achieves the measured relic density either by the $s$-wave dominated process $\tilde{\chi}_1^0 \tilde{\chi}_1^0 \to h_s A_s$ or by the $p$-wave dominated process $\tilde{\chi}_1^0 \tilde{\chi}_1^0 \to h_s h_s$. Since the annihilation cross-section of the s-wave process does not depend on DM velocity at leading order, while that of the p-wave process is proportional to $v^2$, different values of $\kappa$, namely $|\kappa| \sim 0.25$ and $|\kappa| \sim 0.45$, respectively, are needed for the density.  In addition, a small $\lambda$ around 0.01 ensures the suppression of the DM-nucleon scattering, and $v_s \lesssim 1~{\rm TeV}$ is compatible with natural electroweak symmetry breaking.
\item Depending on $M_1$ and $M_2$, the higgsino-dominated $\tilde{\chi}_2^0$ and $\tilde{\chi}_1^\pm$ may be approximately degenerate in mass~\cite{Cao:2021lmj}. In this case, these particles will decay directly into $\tilde{\chi}_1^0$ since there are no other lighter sparticles. On the other hand, the mass splitting of $\tilde{\chi}_2^0$ and $\tilde{\chi}_3^0$ may reach $10~{\rm GeV}$. If the decay $\tilde{\chi}_3^0 \to \tilde{\chi}_1^0 Z$ is kinematically forbidden, the branching ratios of $\tilde{\chi}_3^0 \to \tilde{\chi}_2^0 Z^\ast \to \tilde{\chi}_2^0 f \bar{f}$ and $\tilde{\chi}_3^0 \to \tilde{\chi}_1^0 Z^\ast \to \tilde{\chi}_1^0 f \bar{f}$ are comparable since they are all three-body decays. While if $\tilde{\chi}_3^0 \to \tilde{\chi}_1^0 Z$ is kinematically accessible, the branching ratio of $\tilde{\chi}_3^0 \to \tilde{\chi}_2^0 Z^\ast \to \tilde{\chi}_2^0 f \bar{f}$ is usually negligibly small.
\item For the other heavy sparticles, they tend to decay first into the lighter sparticles other than $\tilde{\chi}_1^0$ given the singlet nature of $\tilde{\chi}_1^0$.  As a result, their decay chain is rather complex.
\item A massive $\tilde{\chi}_1^0$ is utilized to suppress the $E_T^{\rm miss}$ in SUSY signals, and at the same time, massive $\tilde{\mu}_{L,R}$ are adopted to suppress slepton production rates. These mechanisms are helpful for the points to keep consistent with the LHC data.
\end{itemize}

The four points differ mainly in how to  further suppress the ratio $R$. Specifically, points P1 and P2 predict $\tan \beta < 30$ so that $M_2$ should be moderately small to explain the muon g-2 anomaly. Since the wino-dominated $\tilde{\chi}_4^0$ and $\tilde{\chi}_2^\pm$ are copiously produced at the LHC, their decay into $\tilde{\mu}_{L,R}$ must be kinematically forbidden. Otherwise, the probability of $\tilde{\chi}_4^0$ and $\tilde{\chi}_2^\pm$ decaying into leptonic final states will be enhanced by mediating an on-shell $\tilde{\mu}_{L/R}$, which can significantly increase the R-value. In addition, the compressed mass spectra, $\mu_{tot} - m_{\tilde{\chi}_1^0} < m_Z$, are helpful to suppress the signals of the higgsino-dominated electroweakinos. By contrast, points P3 and P4 correspond to $\tan \beta > 50$, and $M_2$ may be exceedingly large in explaining the muon g-2 anomaly. Given the productions of the wino-dominated electroweakinos are suppressed by their large masses, they can not contribute to $R$ significantly even when their decays into $\tilde{\mu}_{L,R}$ are open.

Finally, we add that the samples marked by the sky blue stars in Fig.~\ref{fig1} will be explored at future colliders given that they predict some moderately light sparticles, in particular $m_{\tilde{\chi}_1^0}, m_{\tilde{\chi}_2^0}, m_{\tilde{\chi}_1^\pm} \lesssim 700~{\rm GeV}$. This issue was discussed in Refs.~\cite{Chakraborti:2020vjp,Chakraborti:2021mbr,Chakraborti:2021squ,Chakraborti:2021dli}. It was found that, although only a part of the preferred parameter space can be covered at the high luminosity LHC, an exhaustive coverage of the parameter space is possible at a high-energy $e^+ e^-$ collider with $\sqrt{s} \gtrsim 1~{\rm TeV}$, such as ILC with $\sqrt{s} = 1~{\rm TeV}$~\cite{Baer:2013cma} and CLIC with $\sqrt{s} = 1~{\rm TeV}$~\cite{CLICDetector:2013tfe,CLICdp:2018cto}. This conclusion was shown in Fig.4 of~\cite{Chakraborti:2021squ}, where the capability of different colliders to probe the explanation of the muon g-2 anomaly was compared for the bino-wino co-annihilation case.

\subsection{More discussions of the results}

We explain more about the results of this work.
\begin{enumerate}
\item As suggested by the recent lattice simulation of the BMW collaboration on the hadronic
vacuum polarization contribution to $a_\mu$~\cite{Borsanyi:2020mff}, the muon g-2 anomaly may arise from the uncertainties in calculating the hadronic contribution to the moment. If this speculation is corroborated in the future, $a_\mu^{\rm SUSY}$ should be much smaller than its currently favored size, and any of the electroweakinos and $\tilde{\mu}_{L/R}$ are not necessarily light. In this case, the LHC restrictions on sparticle mass spectra will be relaxed significantly. For example, we updated the results of Ref.~\cite{Cao:2021ljw}, which only studied the DM physics in GNMSSM, by including the recent LZ constraints. We found that the SUSY searches in Table~\ref{Table1} only excluded about $4\%$ of the remaining samples in Fig. 2 of Ref.~\cite{Cao:2021ljw}.

\item As introduced in Section 2.4, the search for charginos and neutralinos using fully hadronic final states of W/Z and Higgs bosons can be more potent in excluding SUSY parameter points than the leptonic signal analysis included in this study. It occurs when the wino-like particles are heavier than about 600 GeV, their mass splitting with $\tilde{\chi}_1^0$ is larger than about 300 GeV, and $m_{\tilde{\chi}_1^0} \lesssim 400~{\rm GeV}$. We study the sky blue samples in the right panel of Fig.~\ref{fig4} that satisfy these conditions and have the following observations:
    \begin{itemize}
    \item Winos are heavier than higgsinos alone for about $4\%$ of the samples, and they are heavier than both higgsinos and $\tilde{\mu}_R$ for another $8\%$ of the samples. Since mass splittings between the winos and the higgsinos are always larger than 100 GeV, the branching ratios of winos decaying into higgsinos are characterized by $Br (\tilde{W} \to \tilde{H} W^\pm ) \gtrsim 50\%$,  $Br (\tilde{W} \to \tilde{H} Z ) < 25\%$, and $Br (\tilde{W} \to \tilde{H} h ) < 25 \%$ ($\tilde{W}$ and $\tilde{H}$ denote the wino-dominated particles and the higgsino-dominated particles, respectively)\footnote{Note that the wino decay modes in the two cases are similar because the strength of wino couplings to $\tilde{\mu}_R$ is much weaker than that to the higgsinos.}, which is shown in Table~\ref{Table6} for benchmark points P1 and P2. In the higgsino co-annihilation case, the higgsinos are approximately degenerate with the singlino-dominated DM in mass and are regarded as missing momentum in the hadronic analysis~\cite{ATLAS:2021yqv}. In this case, one can estimate the R-values. Assuming $Br (\tilde{W} \to \tilde{H} W^\pm ) = 50\%$,  $Br (\tilde{W} \to \tilde{H} Z ) = 25\%$, and $Br (\tilde{W} \to \tilde{H} h ) = 25\%$, and noting that the chargino pair production rate at the LHC is about one-half of the chargino-neutralino associated production rate, we conclude that the R-values of our case are acquired by rescaling those in Fig. 15 of  Ref.~\cite{ATLAS:2021yqv}, with a factor of 0.75, 0.375, and 0.375 for the $W^\pm W^\mp$, $W^\pm Z$, and $W^\pm h$ signal, respectively. In addition, as suggested by the results of the hadronic analysis~\cite{ATLAS:2021yqv}, the mass splitting between the winos and the higgsinos should be larger than about 400 GeV to boost the decay products $W/Z$ and $h$, which can facilitate their detection  using large-radius jets and jet substructure information. We utilize this criterion to refine the two types of samples and find their total percentage is reduced from $12\%$ to $0.7 \%$, indicating that the hadronic analysis may have exclusion capability only for a small portion of the samples. We emphasize that the rescalings further reduce this capability.

    \item Regarding the rest $88\%$ of the samples, we find that winos are always heavier than higgsinos and $\tilde{\mu}_L$. In this case, the wino-like particles may decay into the left-handed dominated sleptons. The branching ratios exceed $20\%$ and may reach $60\%$ in the optimum case. This characteristic is shown in Table~\ref{Table7} for benchmark points P3 and P4. As a result, the branching ratios of the decays into W/Z and $h$ are suppressed as the new channels open up, which can further reduce the capability of the hadronic signals in the SUSY search.
    \end{itemize}

\begin{table}[tpb]
	\centering
	\caption{\label{NLSP}  Numbers of the royal blue samples in the right panel of Fig.~\ref{fig4}, categorized by the NLSP's dominant component and whether the LHC restrictions are considered, are presented in the first three columns of the table, respectively. In the fourth column, we list the SRs that contribute to the largest R-values and their capability to exclude the scenarios in terms of the percentage of the total numbers in the second column.
This table indicates that the LHC restriction on the wino-dominated NLSP scenario is relatively weak, while that on the $\tilde{\mu}_R$ NLSP scenario is very strong. We explained this phenomenon in Ref.~\cite{Cao:2021tuh,Cao:2022chy}. This table also indicates that the experimental analyses in Ref.~\cite{CMS:2017moi} are usually most potent in excluding the samples. }

	\vspace{0.2cm}
	
	\resizebox{0.95\textwidth}{!}{
		\begin{tabular}{l|c|c|l}
			\hline
			NLSP Type & W/O LHC Cons.&  W/ LHC Cons.  & \multicolumn{1}{c}{SRs and their exclusion percentages}     \\ \hline
			\multicolumn{1}{l|}{Total}                                                &9694&4166 & SR-1($24.5\%$), SR-2($14.3\%$), SR-3($6.8\%$), SR-4($5.1\%$)\\
			\multicolumn{1}{l|}{Higgsino}                                      &5003                 &2934 & SR-1($25.4\%$), SR-2($5.3\%$), SR-4($4.5\%$), SR-5($2.5\%$)\\
			\multicolumn{1}{l|}{$\tilde{\nu}_\mu$}                                      &2147              &174 &SR-2($39.6\%$), SR-3($26.7\%$), SR-1($8.3\%$), SR-5($8.3\%$) \\
			\multicolumn{1}{l|}{Wino}                                                                 &1246              &951 &SR-1($9.4\%$), SR-2($8.9\%$), SR-5($2.6\%$), SR-6($1.9\%$) \\
			\multicolumn{1}{l|}{$\tilde{\mu}_{\rm R}$}
			&1013              &18&SR-1($62.7\%$), SR-2($17.6\%$), SR-4($8.8\%$), SR-3($4.9\%$)\\
			Bino                                                                 &285              &89 &SR-1($55.2\%$), SR-2($6.2\%$), SR-4($4.1\%$), SR-5($2.1\%$) \\\hline
	\end{tabular}}
\end{table}

We note that the latest version of the SModelS, namely SModelS-2.2.1~\cite{Alguero:2021dig}, has included the cut efficiency of the hadronic analysis, and it contains some signal topologies that can be applied to some samples of this study. We utilize this code to further refine the sky blue samples in the right panel of Fig.~\ref{fig4}. We find that it has no exclusion capability. We plan to supplement the hadronic analysis into the package CheckMATE to carefully study its impact on SUSY parameter space in our future research project.

Based on the discussions, we conclude that the hadronic analysis is not crucial to the heavy wino case of this study because the decay branching ratios into Z and $h$ are suppressed significantly and also because the momentums of W/Z and $h$ tend to be soft in the cascade decay of the wino-like particles when compared with the scenario that the particles directly decay into $\tilde{\chi}_1^0$. We add that these mechanisms to suppress the capability of the hadronic signal in SUSY search also work in the heavy higgsino case. Furthermore, this case has another distinct characteristic: since winos are usually significantly lighter than 600 GeV when $\mu_{tot} \gtrsim 600~{\rm GeV}$ (see the upper right panel of Fig.~\ref{fig2}), wino production rate is much larger than the higgsino rate, which is less than $10~{\rm fb}$ for $\mu_{tot} = 600~{\rm GeV}$~\cite{Fuks:2012qx,Fuks:2013vua}. In such a situation, it is evident that the leptonic analysis is more favored to exclude the samples.

\item As introduced before, heavy sparticle prefers to decay first into lighter ones other than the LSP. Its decay chain is rather complex, and its final decay products usually contain more than one SM particle. To simplify our discussion of the LHC restrictions,  we classify the royal blue samples in the right panel of Fig.~\ref{fig4} by their NLSP’s dominant component, which may be any of higgsino fields, wino field, bino field, left-handed slepton field, and right-handed slepton field. Given that the critical characteristics of these scenarios and their underlying physics have been analyzed in our recent works~\cite{Cao:2021tuh,Cao:2022chy}, we only reveal more details about the exclusion capability of the SUSY searches at the LHC. In Table~\ref{NLSP}, we show sample numbers of these scenarios before and after considering the LHC restrictions, the SRs that contribute to the largest R-values, and their capability to exclude the scenarios, expressed in terms of the percentage of the total numbers in the second column. The following SRs are involved:
\begin{itemize}
	\item SR-1: Signal regions $\rm{G03}$, $\rm{G04}$, and $\rm{G05}$ defined in Ref.~\cite{CMS:2017moi}. They correspond to the LHC search for electroweakinos which focused on the events with $p_{\rm T}^{\rm miss} > 100~{\rm GeV}$ ($p_{\rm T}^{\rm miss}$ denotes missing transverse momentum) and four or more leptons that form at least two opposite-sign same-flavor (OSSF) pairs.
	\item SR-2: Signal regions $\rm{A44}$, $\rm{A14}$, $\rm{A30}$, and so on defined in Ref.~\cite{CMS:2017moi}. They arise from the LHC search for electroweakinos by the final state containing missing transverse momentum and three electrons or muons that form at least one OSSF pair.
	\item SR-3: Signal regions $\rm{SS15}$, $\rm{SS14}$, and $\rm{SS12}$ in Ref.~\cite{CMS:2017moi}. They are designed in the LHC search for electroweakinos by the final state containing two same-sign (SS) dileptons with $p_{\rm T} > 100~{\rm GeV}$ and $M_{\rm T} > 100~{\rm GeV}$, and no jets.
	\item SR-4: Signal regions $\rm{SR\_WZoff\_high\_njd}$, $\rm{SR\_incWZoff\_high\_njc1}$, $\rm{SR\_WZ}$ $\rm{off\_high\_nje}$, and so on defined in Ref.~\cite{ATLAS:2021moa}. They study the chargino-neutralino associated production at the LHC by the final state containing three leptons and missing transverse momentum, where the chargino and neutralino decay into off-shell W and Z bosons, respectively.
	\item SR-5: Signal regions $\rm{S-high-mm-10}$, $\rm{S-high-mm-05}$, $\rm{E-high}$ $\rm{-mm-30}$, and so on proposed in Ref.~\cite{ATLAS:2019lng}. They consider the electroweakinos with compressed mass spectra
and investigate their production at the LHC by the final state containing two leptons and missing transverse momentum.
	\item SR-6: Signal regions $\rm{SR\_Wh\_SFOS\_11}$, $\rm{ SR\_WZ\_20}$, $\rm{SR\_Wh\_SFOS\_16}$, and so on defined in Ref.~\cite{ATLAS:2021moa}. They correspond to the LHC search for chargino-neutralino associated production by the final state containing three leptons and missing transverse momentum, where the chargino decays into an on-shell W boson, and the neutralino decays into an on-shell Z boson or an on-shell SM Higgs boson.
\end{itemize}

In rare cases, the following SRs are also crucial:
\begin{itemize}
	\item SR-7: Signal regions $\rm{SRG08\_0j\_mll}$ and $\rm{SRG07\_0j\_mll}$ defined in Ref.~\cite{CMS:2020bfa}. They come from the LHC search for new physics beyond the SM, focusing on the final state containing two OSSF leptons, jets, and missing transverse momentum.
	\item SR-8: Signal regions $\rm{SR1\_weakino\_2media\_mll\_2}$ and $\rm{SR1\_weakino\_3high\_}$ ${\rm mll\_2}$ defined in Ref.~\cite{CMS:2018szt}. They arise from the LHC search for new physics by the signal containing two soft oppositely charged leptons and missing transverse momentum.
\end{itemize}
We add that the details of these SRs were presented in corresponding experimental reports. They are helpful to understand the results in Table~\ref{NLSP}.

\end{enumerate}
			
\section{\label{conclusion-section}Summary}

In the past decade, DM direct detection experiments have improved their sensitivity to the cross-sections of DM-nucleon scattering by more than $10^3$ times~\cite{Aprile:2018dbl,Aprile:2019dbj,LZ:2022ufs,PandaX-4T:2021bab}. Consequently, some economical realizations of SUSY, such as the MSSM and the $Z_3$-NMSSM, are becoming more and more difficult to keep consistent with the experimental results in a natural way~\cite{Cao:2019qng}. In particular, the attractiveness of the popular bino-dominated DM in these models is fading. In this context, the singlino-dominated DM in the GNMSSM arouses the researchers' interest~\cite{Cao:2021ljw}. This theory has the following distinct features~\cite{Cao:2021tuh}: free from the tadpole problem and the domain-wall problem of the $Z_3$-NMSSM, capable of forming an economical secluded DM sector to yield the DM experimental results naturally, and readily weaken the restrictions from the LHC search for SUSY. In addition, it predicts more stable vacuums than the $Z_3$-NMSSM. As a result, the theory can explain the muon g-2 anomaly in broad parameter space that agrees with all experimental results while simultaneously breaking the electroweak symmetry naturally.

In this study, we are inspired by the rapid progress of particle physics experiments in recent years and investigate how the GNMSSM coincides with various experimental data. We are particularly interested in the SUSY search at the LHC, the DM search by the LZ experiment, and the improved measurement of the muon g-2 since they can provide valuable information about SUSY. In order to survey the theory's status, we employ the MultiNest algorithm to scan elaborately its parameter space. We adopt the muon g-2 observable to guide the scan and consider the restrictions from the LHC Higgs data, the DM experimental results, the B-physics measurements, and the vacuum stability. Then, we examine
the samples obtained from the scan by the LHC analyses in sparticle search. Our study reveals three spectacular features of the theory. The first is that there exist lower bounds on sparticle mass spectra, e.g., $\mu_{tot} \gtrsim 210~{\rm GeV}$, $M_2 \gtrsim 260~{\rm GeV}$, $m_{\tilde{\chi}_1^0} \gtrsim 140~{\rm GeV}$,  $m_{\tilde{\chi}_2^0}, m_{\tilde{\chi}_1^\pm} \gtrsim 200~{\rm GeV}$,  $m_{\tilde{\chi}_3^0} \gtrsim 250~{\rm GeV}$, $m_{\tilde{\mu}_L} \gtrsim 255~{\rm GeV}$, and $m_{\tilde{\mu}_R} \gtrsim 240~{\rm GeV}$. These bounds are significantly lower than those for the bino-dominated DM case in the $Z_3$-NMSSM~\cite{Cao:2022htd}, but far beyond the reach of the LEP experiments in searching for SUSY. They originate from the following facts: if $\tilde{\chi}_1^0$ is lighter, more missing transverse energy will be emitted in the sparticle production processes at the LHC, which can improve the sensitivities of the experimental analyses; while if the sparticles other than $\tilde{\chi}_1^0$ are lighter, they will be more copiously produced at the LHC to increase the events containing multiple leptons. The second feature is that the singlet-dominated $\tilde{\chi}_1^0$, $h_s$, and $A_s$ form a secluded DM sector. Such a theoretical structure is natural in the sense that the masses of these particles can be regarded as free parameters, and all of them and $v_s$ are at the weak scale. The relic density can be obtained either by the $s$-wave dominated annihilation $\tilde{\chi}_1^0 \tilde{\chi}_1^0 \to h_s A_s$ with $|\kappa| \sim 0.25$ or by the $p$-wave dominated annihilation $\tilde{\chi}_1^0 \tilde{\chi}_1^0 \to h_s h_s$ with $|\kappa| \sim 0.45$. The last feature is that the improving detections of the DM-nucleon scattering influence the theory only by preferring a smaller and smaller $\lambda$, which is currently upper bounded by about $0.05$. Since the `visible' sector (in comparison with the DM sector) is scarcely affected, one does not need to worry about the drastic change of the theory's phenomenology if the scattering rate is found below the neutrino floor.  Evidently, our study provides a simple and clear picture of the physics inherent in the GNMSSM, although it possesses more input parameters than the MSSM and the $Z_3$-NMSSM.

This work extends the study in~\cite{Cao:2018rix} by considering a more general theoretical framework and utilizing more advanced and sophisticated research strategies. As a result, the conclusions obtained in this work are more robust than those of the previous work. They exhibit the most essential characteristics of the NMSSM.

\section*{Acknowledgement}
This work is supported by the National Natural Science Foundation of China (NNSFC) under grant No. 12075076.
		
\bibliographystyle{CitationStyle}
\bibliography{g-2}

\end{document}